\documentclass[journal,draftclsnofoot,onecolumn,12pt]{IEEEtran}

\usepackage{graphicx}
\usepackage{subfigure}
\usepackage{multirow}
\usepackage{array}
\newcolumntype{P}[1]{>{\centering\arraybackslash}p{#1}}
\newcolumntype{M}[1]{>{\centering\arraybackslash}m{#1}}

\usepackage{amsthm,amssymb,amsmath,graphicx,multirow,color,amsfonts}%
\usepackage[update,prepend]{epstopdf}
\usepackage{multirow}
\usepackage[latin1]{inputenc}
\usepackage{tikz}
\usepackage{bbm} 
\usepackage{pdfpages}
\usepackage{multirow}
\usepackage{subfig}
\usepackage{comment}

\usepackage{graphicx}
\usepackage{multicol}
\usepackage{cite}

\usepackage[justification=centering]{caption}
\usepackage{textcomp}
\usepackage{psfrag}
\usepackage{arydshln}
\usepackage{url}
\usepackage{soul}
\usepackage{graphicx,color}
\usepackage[nolist]{acronym}
\usepackage{algorithm,algorithmic} 

\usepackage{mathtools,lipsum}
\usepackage{cuted}
\usepackage{amsmath}
\newtheorem{proposition}{Proposition}
\usepackage{mathrsfs}

\usepackage[capitalise]{cleveref}
\Crefname{equation}{Eq.\!}{Eqs.\!}
\Crefname{figure}{Fig.\!}{Figs.\!}
\Crefname{tabular}{Tab.\!}{Tabs.\!}
\Crefname{section}{Section\!}{Sections.\!}

\def\nb0{{\mathbf{0}}}
\def\nb1{{\mathbf{1}}}

\newtheorem{lemma}{Lemma}

\newtheorem{definition}{Definition}

\newtheorem{theorem}{Theorem}

\def\argmin{\operatorname{arg~min}}

\newenvironment{sequation}{
\begin{equation}\small}{\end{equation}
}

\begin{document}
\graphicspath{{./Figures/}}
	\begin{acronym}

\acro{5G-NR}{5G New Radio}
\acro{3GPP}{3rd Generation Partnership Project}
\acro{ABS}{aerial base station}
\acro{AC}{address coding}
\acro{ACF}{autocorrelation function}
\acro{ACR}{autocorrelation receiver}
\acro{ADC}{analog-to-digital converter}
\acrodef{aic}[AIC]{Analog-to-Information Converter}     
\acro{AIC}[AIC]{Akaike information criterion}
\acro{aric}[ARIC]{asymmetric restricted isometry constant}
\acro{arip}[ARIP]{asymmetric restricted isometry property}

\acro{ARQ}{Automatic Repeat Request}
\acro{AUB}{asymptotic union bound}
\acrodef{awgn}[AWGN]{Additive White Gaussian Noise}     
\acro{AWGN}{additive white Gaussian noise}

\acro{APSK}[PSK]{asymmetric PSK} 

\acro{waric}[AWRICs]{asymmetric weak restricted isometry constants}
\acro{warip}[AWRIP]{asymmetric weak restricted isometry property}
\acro{BCH}{Bose, Chaudhuri, and Hocquenghem}        
\acro{BCHC}[BCHSC]{BCH based source coding}
\acro{BEP}{bit error probability}
\acro{BFC}{block fading channel}
\acro{BG}[BG]{Bernoulli-Gaussian}
\acro{BGG}{Bernoulli-Generalized Gaussian}
\acro{BPAM}{binary pulse amplitude modulation}
\acro{BPDN}{Basis Pursuit Denoising}
\acro{BPPM}{binary pulse position modulation}
\acro{BPSK}{Binary Phase Shift Keying}
\acro{BPZF}{bandpass zonal filter}
\acro{BSC}{binary symmetric channels}              
\acro{BU}[BU]{Bernoulli-uniform}
\acro{BER}{bit error rate}
\acro{BS}{base station}
\acro{BW}{BandWidth}
\acro{BLLL}{ binary log-linear learning }

\acro{CP}{Cyclic Prefix}
\acrodef{cdf}[CDF]{cumulative distribution function}   
\acro{CDF}{Cumulative Distribution Function}
\acrodef{c.d.f.}[CDF]{cumulative distribution function}
\acro{CCDF}{complementary cumulative distribution function}
\acrodef{ccdf}[CCDF]{complementary CDF}               
\acrodef{c.c.d.f.}[CCDF]{complementary cumulative distribution function}
\acro{CD}{cooperative diversity}

\acro{CDMA}{Code Division Multiple Access}
\acro{ch.f.}{characteristic function}
\acro{CIR}{channel impulse response}
\acro{cosamp}[CoSaMP]{compressive sampling matching pursuit}
\acro{CR}{cognitive radio}
\acro{cs}[CS]{compressed sensing}                   
\acrodef{cscapital}[CS]{Compressed sensing} 
\acrodef{CS}[CS]{compressed sensing}
\acro{CSI}{channel state information}
\acro{CCSDS}{consultative committee for space data systems}
\acro{CC}{convolutional coding}
\acro{Covid19}[COVID-19]{Coronavirus disease}

\acro{DAA}{detect and avoid}
\acro{DAB}{digital audio broadcasting}
\acro{DCT}{discrete cosine transform}
\acro{dft}[DFT]{discrete Fourier transform}
\acro{DR}{distortion-rate}
\acro{DS}{direct sequence}
\acro{DS-SS}{direct-sequence spread-spectrum}
\acro{DTR}{differential transmitted-reference}
\acro{DVB-H}{digital video broadcasting\,--\,handheld}
\acro{DVB-T}{digital video broadcasting\,--\,terrestrial}
\acro{DL}{DownLink}
\acro{DSSS}{Direct Sequence Spread Spectrum}
\acro{DFT-s-OFDM}{Discrete Fourier Transform-spread-Orthogonal Frequency Division Multiplexing}
\acro{DAS}{Distributed Antenna System}
\acro{DNA}{DeoxyriboNucleic Acid}

\acro{EC}{European Commission}
\acro{EED}[EED]{exact eigenvalues distribution}
\acro{EIRP}{Equivalent Isotropically Radiated Power}
\acro{ELP}{equivalent low-pass}
\acro{eMBB}{Enhanced Mobile Broadband}
\acro{EMF}{ElectroMagnetic Field}
\acro{EU}{European union}
\acro{EI}{Exposure Index}
\acro{eICIC}{enhanced Inter-Cell Interference Coordination}

\acro{FC}[FC]{fusion center}
\acro{FCC}{Federal Communications Commission}
\acro{FEC}{forward error correction}
\acro{FFT}{fast Fourier transform}
\acro{FH}{frequency-hopping}
\acro{FH-SS}{frequency-hopping spread-spectrum}
\acrodef{FS}{Frame synchronization}
\acro{FSsmall}[FS]{frame synchronization}  
\acro{FDMA}{Frequency Division Multiple Access}

\acro{GA}{Gaussian approximation}
\acro{GF}{Galois field }
\acro{GG}{Generalized-Gaussian}
\acro{GIC}[GIC]{generalized information criterion}
\acro{GLRT}{generalized likelihood ratio test}
\acro{GPS}{Global Positioning System}
\acro{GMSK}{Gaussian Minimum Shift Keying}
\acro{GSMA}{Global System for Mobile communications Association}
\acro{GS}{ground station}
\acro{GMG}{ Grid-connected MicroGeneration}

\acro{HAP}{high altitude platform}
\acro{HetNet}{Heterogeneous network}

\acro{IDR}{information distortion-rate}
\acro{IFFT}{inverse fast Fourier transform}
\acro{iht}[IHT]{iterative hard thresholding}
\acro{i.i.d.}{independent, identically distributed}
\acro{IoT}{Internet of Things}                      
\acro{IR}{impulse radio}
\acro{lric}[LRIC]{lower restricted isometry constant}
\acro{lrict}[LRICt]{lower restricted isometry constant threshold}
\acro{ISI}{intersymbol interference}
\acro{ITU}{International Telecommunication Union}
\acro{ICNIRP}{International Commission on Non-Ionizing Radiation Protection}
\acro{IEEE}{Institute of Electrical and Electronics Engineers}
\acro{ICES}{IEEE international committee on electromagnetic safety}
\acro{IEC}{International Electrotechnical Commission}
\acro{IARC}{International Agency on Research on Cancer}
\acro{IS-95}{Interim Standard 95}

\acro{KPI}{Key Performance Indicator}

\acro{LEO}{low earth orbit}
\acro{LF}{likelihood function}
\acro{LLF}{log-likelihood function}
\acro{LLR}{log-likelihood ratio}
\acro{LLRT}{log-likelihood ratio test}
\acro{LoS}{Line-of-Sight}
\acro{LRT}{likelihood ratio test}
\acro{wlric}[LWRIC]{lower weak restricted isometry constant}
\acro{wlrict}[LWRICt]{LWRIC threshold}
\acro{LPWAN}{Low Power Wide Area Network}
\acro{LoRaWAN}{Low power long Range Wide Area Network}
\acro{NLoS}{Non-Line-of-Sight}
\acro{LiFi}[Li-Fi]{light-fidelity}
 \acro{LED}{light emitting diode}
 \acro{LABS}{LoS transmission with each ABS}
 \acro{NLABS}{NLoS transmission with each ABS}

\acro{MB}{multiband}
\acro{MC}{macro cell}
\acro{MDS}{mixed distributed source}
\acro{MF}{matched filter}
\acro{m.g.f.}{moment generating function}
\acro{MI}{mutual information}
\acro{MIMO}{Multiple-Input Multiple-Output}
\acro{MISO}{multiple-input single-output}
\acrodef{maxs}[MJSO]{maximum joint support cardinality}                       
\acro{ML}[ML]{maximum likelihood}
\acro{MMSE}{minimum mean-square error}
\acro{MMV}{multiple measurement vectors}
\acrodef{MOS}{model order selection}
\acro{M-PSK}[${M}$-PSK]{$M$-ary phase shift keying}                       
\acro{M-APSK}[${M}$-PSK]{$M$-ary asymmetric PSK} 
\acro{MP}{ multi-period}
\acro{MINLP}{mixed integer non-linear programming}

\acro{M-QAM}[$M$-QAM]{$M$-ary quadrature amplitude modulation}
\acro{MRC}{maximal ratio combiner}                  
\acro{maxs}[MSO]{maximum sparsity order}                                      
\acro{M2M}{Machine-to-Machine}                                                
\acro{MUI}{multi-user interference}
\acro{mMTC}{massive Machine Type Communications}      
\acro{mm-Wave}{millimeter-wave}
\acro{MP}{mobile phone}
\acro{MPE}{maximum permissible exposure}
\acro{MAC}{media access control}
\acro{NB}{narrowband}
\acro{NBI}{narrowband interference}
\acro{NLA}{nonlinear sparse approximation}
\acro{NLOS}{Non-Line of Sight}
\acro{NTIA}{National Telecommunications and Information Administration}
\acro{NTP}{National Toxicology Program}
\acro{NHS}{National Health Service}

\acro{LOS}{Line of Sight}

\acro{OC}{optimum combining}                             
\acro{OC}{optimum combining}
\acro{ODE}{operational distortion-energy}
\acro{ODR}{operational distortion-rate}
\acro{OFDM}{Orthogonal Frequency-Division Multiplexing}
\acro{omp}[OMP]{orthogonal matching pursuit}
\acro{OSMP}[OSMP]{orthogonal subspace matching pursuit}
\acro{OQAM}{offset quadrature amplitude modulation}
\acro{OQPSK}{offset QPSK}
\acro{OFDMA}{Orthogonal Frequency-division Multiple Access}
\acro{OPEX}{Operating Expenditures}
\acro{OQPSK/PM}{OQPSK with phase modulation}

\acro{PAM}{pulse amplitude modulation}
\acro{PAR}{peak-to-average ratio}
\acrodef{pdf}[PDF]{probability density function}                      
\acro{PDF}{probability density function}
\acrodef{p.d.f.}[PDF]{probability distribution function}
\acro{PDP}{power dispersion profile}
\acro{PMF}{probability mass function}                             
\acrodef{p.m.f.}[PMF]{probability mass function}
\acro{PN}{pseudo-noise}
\acro{PPM}{pulse position modulation}
\acro{PRake}{Partial Rake}
\acro{PSD}{power spectral density}
\acro{PSEP}{pairwise synchronization error probability}
\acro{PSK}{phase shift keying}
\acro{PD}{power density}
\acro{8-PSK}[$8$-PSK]{$8$-phase shift keying}
\acro{PPP}{Poisson point process}
\acro{PCP}{Poisson cluster process}
 
\acro{FSK}{Frequency Shift Keying}

\acro{QAM}{Quadrature Amplitude Modulation}
\acro{QPSK}{Quadrature Phase Shift Keying}
\acro{OQPSK/PM}{OQPSK with phase modulator }

\acro{RD}[RD]{raw data}
\acro{RDL}{"random data limit"}
\acro{ric}[RIC]{restricted isometry constant}
\acro{rict}[RICt]{restricted isometry constant threshold}
\acro{rip}[RIP]{restricted isometry property}
\acro{ROC}{receiver operating characteristic}
\acro{rq}[RQ]{Raleigh quotient}
\acro{RS}[RS]{Reed-Solomon}
\acro{RSC}[RSSC]{RS based source coding}
\acro{r.v.}{random variable}                               
\acro{R.V.}{random vector}
\acro{RMS}{root mean square}
\acro{RFR}{radiofrequency radiation}
\acro{RIS}{Reconfigurable Intelligent Surface}
\acro{RNA}{RiboNucleic Acid}
\acro{RRM}{Radio Resource Management}
\acro{RUE}{reference user equipments}
\acro{RAT}{radio access technology}
\acro{RB}{resource block}

\acro{SA}[SA-Music]{subspace-augmented MUSIC with OSMP}
\acro{SC}{small cell}
\acro{SCBSES}[SCBSES]{Source Compression Based Syndrome Encoding Scheme}
\acro{SCM}{sample covariance matrix}
\acro{SEP}{symbol error probability}
\acro{SG}[SG]{sparse-land Gaussian model}
\acro{SIMO}{single-input multiple-output}
\acro{SINR}{signal-to-interference plus noise ratio}
\acro{SIR}{signal-to-interference ratio}
\acro{SISO}{Single-Input Single-Output}
\acro{SMV}{single measurement vector}
\acro{SNR}[\textrm{SNR}]{signal-to-noise ratio} 
\acro{sp}[SP]{subspace pursuit}
\acro{SS}{spread spectrum}
\acro{SW}{sync word}
\acro{SAR}{specific absorption rate}
\acro{SSB}{synchronization signal block}
\acro{SR}{shrink and realign}

\acro{tUAV}{tethered Unmanned Aerial Vehicle}
\acro{TBS}{terrestrial base station}

\acro{uUAV}{untethered Unmanned Aerial Vehicle}
\acro{PDF}{probability density functions}

\acro{PL}{path-loss}

\acro{TH}{time-hopping}
\acro{ToA}{time-of-arrival}
\acro{TR}{transmitted-reference}
\acro{TW}{Tracy-Widom}
\acro{TWDT}{TW Distribution Tail}
\acro{TCM}{trellis coded modulation}
\acro{TDD}{Time-Division Duplexing}
\acro{TDMA}{Time Division Multiple Access}
\acro{Tx}{average transmit}

\acro{UAV}{Unmanned Aerial Vehicle}
\acro{uric}[URIC]{upper restricted isometry constant}
\acro{urict}[URICt]{upper restricted isometry constant threshold}
\acro{UWB}{ultrawide band}
\acro{UWBcap}[UWB]{Ultrawide band}   
\acro{URLLC}{Ultra Reliable Low Latency Communications}
         
\acro{wuric}[UWRIC]{upper weak restricted isometry constant}
\acro{wurict}[UWRICt]{UWRIC threshold}                
\acro{UE}{User Equipment}
\acro{UL}{UpLink}

\acro{WiM}[WiM]{weigh-in-motion}
\acro{WLAN}{wireless local area network}
\acro{wm}[WM]{Wishart matrix}                               
\acroplural{wm}[WM]{Wishart matrices}
\acro{WMAN}{wireless metropolitan area network}
\acro{WPAN}{wireless personal area network}
\acro{wric}[WRIC]{weak restricted isometry constant}
\acro{wrict}[WRICt]{weak restricted isometry constant thresholds}
\acro{wrip}[WRIP]{weak restricted isometry property}
\acro{WSN}{wireless sensor network}                        
\acro{WSS}{Wide-Sense Stationary}
\acro{WHO}{World Health Organization}
\acro{Wi-Fi}{Wireless Fidelity}

\acro{sss}[SpaSoSEnc]{sparse source syndrome encoding}

\acro{VLC}{Visible Light Communication}
\acro{VPN}{Virtual Private Network} 
\acro{RF}{Radio Frequency}
\acro{FSO}{Free Space Optics}
\acro{IoST}{Internet of Space Things}

\acro{GSM}{Global System for Mobile Communications}
\acro{2G}{Second-generation cellular network}
\acro{3G}{Third-generation cellular network}
\acro{4G}{Fourth-generation cellular network}
\acro{5G}{Fifth-generation cellular network}	
\acro{gNB}{next-generation Node-B Base Station}
\acro{NR}{New Radio}
\acro{UMTS}{Universal Mobile Telecommunications Service}
\acro{LTE}{Long Term Evolution}

\acro{QoS}{Quality of Service}
\end{acronym}
	
\newcommand{\SAR} {\mathrm{SAR}}
\newcommand{\WBSAR} {\mathrm{SAR}_{\mathsf{WB}}}
\newcommand{\gSAR} {\mathrm{SAR}_{10\si{\gram}}}
\newcommand{\Sab} {S_{\mathsf{ab}}}
\newcommand{\Eavg} {E_{\mathsf{avg}}}
\newcommand{\ft}{f_{\textsf{th}}}
\newcommand{\alphatf}{\alpha_{24}}


\title{
Analyzing Localizability of LEO/MEO Hybrid Networks: A Stochastic Geometry Approach
}

\author{
Ruibo Wang, {\em Member, IEEE}, Mustafa A. Kishk, {\em Member, IEEE}, \\ Howard H. Yang, {\em Member, IEEE}, and Mohamed-Slim Alouini, {\em Fellow, IEEE}
\thanks{Ruibo Wang and Mohamed-Slim Alouini are with King Abdullah University of Science and Technology (KAUST), CEMSE division, Thuwal 23955-6900, Saudi Arabia. Mustafa A. Kishk is with the Department of Electronic Engineering, Maynooth University, Maynooth, W23 F2H6, Ireland. Howard H. Yang is with the ZJU-UIUC Institute, Zhejiang University, Haining 314400, China. (e-mail: ruibo.wang@kaust.edu.sa; mustafa.kishk@mu.ie; haoyang@intl.zju.edu.cn; slim.alouini@kaust.edu.sa). Corresponding author: Howard H. Yang.
}
\vspace{-8mm}
}
\maketitle

\vspace{-0.8cm}

\begin{abstract}
With the increase in global positioning service demands and the requirement for more precise positioning, assisting existing medium and high orbit satellite-enabled positioning systems with low Earth orbit (LEO) satellites has garnered widespread attention. However, providing low computational complexity performance analysis for hybrid LEO/MEO massive satellite constellations remains a challenge. In this article, we introduce for the first time the application of stochastic geometry (SG) framework in satellite-enabled positioning performance analysis and provide an analytical expression for the $K-$availiability probability and $K-$localizability probability under bidirectional beam alignment transmissions. The $K-$localizability probability, defined as the probability that at least $K$ satellites can participate in the positioning process, serves as a prerequisite for positioning. Since the modeling of MEO satellite constellations within the SG framework has not yet been studied, we integrate the advantages of Cox point processes and binomial point processes, proposing a doubly stochastic binomial point process binomial point process for accurate modeling of MEO satellite constellations. Finally, we investigate the impact of constellation configurations and antenna patterns on the localizability performance of LEO, MEO, and hybrid MEO/LEO constellations. We also demonstrate the network performance gains brought to MEO positioning systems by incorporating assistance from LEO satellites. 
\end{abstract}


\section{Introduction}
Over the years, satellite-enabled positioning services have been primarily enabled by medium Earth orbit (MEO) satellites and a small number of geosynchronous Earth orbit (GEO) satellites. 
Indeed, positioning technologies based on MEO has been mature, achieving extensive coverage as well as global positioning with a small number of satellites~\cite{cai2015precise}. However, the already deployed satellite constellations may face challenges in expanding their capabilities when addressing the demands of future networks for low latency and high-precision positioning, as well as the increasing requirements for additional positioning services \cite{ferre2022leo}. Involving the deployment of mega LEO satellite constellations in auxiliary positioning is a potential solution \cite{liao2023integration}. Compared to MEO satellites, LEO satellites orbit at a closer distance to the Earth, resulting in lower positioning latency \cite{yue2022satellite, reid2018broadband}. 
Besides, since the carrier frequencies used by LEO satellites are different from those for MEO satellite-enabled positioning systems~\cite{ferre2021comparison}, the two systems do not interfere with each other while positioning errors can be reduced with an increased number of satellites in use~\cite{specht2015accuracy}.
Therefore, leveraging LEO satellites in existing MEO-based systems to improve positioning performance is a topic worthy of research.

\par
Nonetheless, assessing the performance of such an integration poses new challenges. 
More precisely, the performance of traditional MEO satellite-enabled positioning systems is chiefly evaluated based on simulations. 
Specifically, researchers use simulation software to generate the positions of satellites at a particular moment \cite{raghu2016tracking}. Performance testing of proposed algorithms and analytical results is then conducted based on these simulated satellite positions \cite{halevi2022asymptotic,wei2023time}. To ensure the reliability of the results, the above procedures need to be executed in numerous rounds to validate the performance gain under various satellite topologies \cite{chandrika2023spin}. 
Compared to MEO constellations, analyzing the performance of positioning in a hybrid LEO/MEO satellite system is more difficult. 
Currently, the number of satellites in medium to high Earth orbit global positioning systems is only a few dozen, making simulation-based satellite position modeling reasonable \cite{cai2015precise}. However, the proliferation of mega-constellations of LEO satellites, numbering in the hundreds or even thousands, significantly increases computational complexity \cite{al2022next}. An effective solution is to establish an analytical framework to directly map the satellite system's parameters to positioning-related metrics, such as satellite diversity and accuracy of positioning. Among various mathematical tools, stochastic geometry (SG) is particularly well-suited for establishing the aforementioned analytical framework, as it is also one of the few tools capable of interference analysis in large-scale random network topology \cite{wang2022ultra}.

\subsection{Related Works}\label{section1-2}
In this subsection, we briefly review existing research on SG-based satellite modeling and SG-based positioning for terrestrial networks, emphasizing the foundation upon which this article is built, along with their limitations.

\par
In recent years, there have been studies employing SG for modeling and analyzing LEO satellite networks under the SG framework \cite{Al-2, wang2025modeling}. Among the SG-based models, the spherical binomial point process (BPP) is one of the most widely applied models for LEO satellite constellations. The BPP requires only the altitude of the constellation and the number of satellites, to model the LEO satellite constellation \cite{talgat2020stochastic}. However, compared to real-world constellations, the accuracy of BPP modeling diminishes with a decrease in the number of satellites \cite{wang2022evaluating}. Although the BPP model has been shown to align with real-world constellations in terms of performance evaluation for massive LEO constellations containing thousands of satellites \cite{ok-1}, it is not a suitable modeling approach for MEO constellations that consist of only dozens of satellites.

\par
Furthermore, the Cox point process (CPP) model proposed by the authors in \cite{choi2024novel,choi2024modeling} offers inspiration for modeling MEO satellite constellations. Although the CPP model is still applied to modeling mega LEO constellations, it takes into account the reality of satellites operating on orbits, thus making it more accurate than BPP. CPP assumes that both the number of orbits and the number of satellites per orbit follow a Poisson distribution \cite{choi2024novel}. Authors in \cite{jung2023modeling}, by simplifying the CPP model to a constellation with a single orbit and applying it to GEO satellite modeling, have demonstrated to some extent that orbit-based models are more reasonable and accurate for modeling smaller-scale constellations like MEO. While the performance differences between the Poisson distribution and the binomial distribution become negligible when the number of satellites is large, a MEO constellation typically consists of only several orbits with each orbit hosting several satellites. In such cases, modeling the number of orbits and the number of satellites per orbit as random Poisson distributions deviates from the deterministic reality of satellite constellations, leading to discrepancies between the model and the actual deterministic constellation model. Hence, it is necessary to propose a more accurate SG-based model for MEO constellations with a limited number of satellites.

\par
So far, researchers have established analytical frameworks for positioning in planar networks under the SG framework. Among them, almost all of the studies analyzed the localizability probability, as it is one of the most fundamental and important metrics for localization \cite{schloemann2015localization,christopher2018statistical}. The $K-$localizability probability is defined as the probability that a target can detect at least $K$ signals \cite{schloemann2015toward}. It is worth noting that the definition of "localizability" in satellite-enabled positioning systems differs significantly from that of terrestrial-based positioning systems. Since single-satellite localization will be used in practice, satisfying $K-$localizability ($K>1$) does not necessarily mean that the positioning system requires the detection of at least $K$ satellites to function properly. Considering the mentioned fact that the more satellites are involved in positioning, the better the performance and the higher the accuracy, we have continued to use the concept of localizability in satellite-enabled positioning. In this context, $K-$localizability is used to measure the quality of localization performance rather than the feasibility of the positioning system's operation.

\par
In addition, the analysis of satellite networks also differs significantly from that of terrestrial networks. Satellites are deployed on a spherical surface, making the analytical expression of distance distribution more complicated \cite{talgat2020nearest} than the planar network. Satellites are equipped with antennas emitting directional signals, making interference analysis challenging \cite{okati2022nonhomogeneous}. Furthermore, only satellites not blocked by the Earth can participate in positing, making satellite availability a prerequisite for localizability \cite{wang2022stochastic}.

\subsection{Contribution}\label{section1-3}
To our knowledge, this is the first article that applies the SG framework in the localizability analysis of satellite networks. The contributions of this article are as follows.
\begin{itemize}
    \item This is the first study of MEO satellite constellations modeling and analysis under the SG framework. We propose a doubly stochastic binomial point process (DSBPP) for modeling MEO satellites. Specifically, an underlying BPP is generated and mapped to the positions of orbits, with satellites on each orbit following a BPP distribution based on the underlying BPP. Compared to CPP, DSBPP provides a more accurate modeling of MEO constellations without decreasing the tractability.
    \item For interference mitigation considerations, we adopt a bi-directional aligned beam model and provide a general representation for modeling antenna gains. Because existing literature on the spherical SG framework primarily discusses unidirectional beam alignment, our analytical framework can offer guidance for subsequent bidirectional alignment.
    \item Based on this model, we derive analytical expressions for availability and estimate the localizability of satellite-enabled positioning systems under the precondition of satellite availability. The above analytical results are of low complexity and verified to be accurate.
    \item Finally, the impacts of constellation configurations such as constellation altitude, and physical-level parameters such as antenna patterns on network performance are studied. In addition, we demonstrate that a hybrid MEO/LEO constellation offers better localizability performance compared to an MEO constellation without the assistance of LEO satellites.
\end{itemize}

\section{System Model}\label{section2}
\begin{figure*}[ht]
\centering
\includegraphics[width=\linewidth]{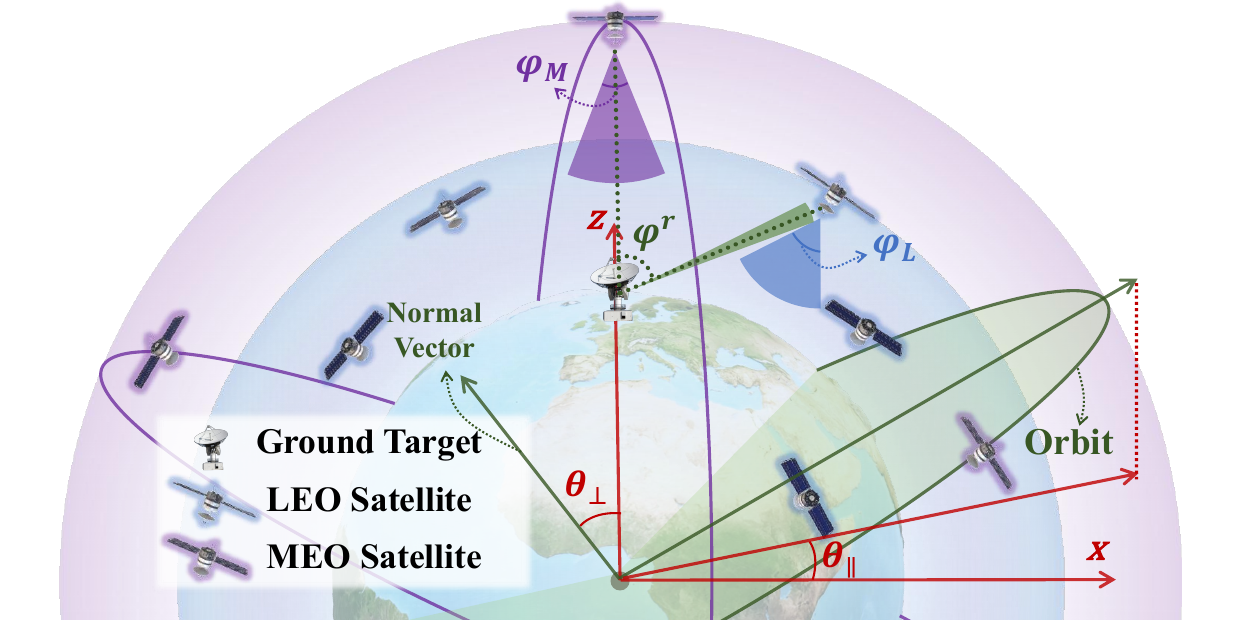}
\caption{Schematic diagram for the system model.}
\label{figure1}
\end{figure*}

This section provides the LEO and MEO satellite spatial distribution models, antenna gain pattern models, and the satellite-terrestrial channel model. A schematic diagram for the system model is shown in Fig.~\ref{figure1}.

\subsection{Spatial Distribution}
We consider a hybrid positioning system composed of an LEO satellite constellation and an MEO satellite constellation. The LEO satellite constellation consists of $N_L$ satellites distributed on a sphere around Earth, forming a homogeneous spherical BPP. The distance between the LEO satellite and Earth's center is denoted as $R_L$.
According to Slivnyak's theorem, homogeneous BPP is rotation invariant \cite{feller1991introduction}. In other words, the rotation of the coordinate system does not affect the LEO satellite distribution. Therefore, we take the Earth's center as the origin. Without loss of generality, the coordinates of a typical ground target in the spherical coordinate system as $(R_{\oplus},0,0)$, where $R_{\oplus}=6371$~km is the radius of the Earth. The azimuth angle and polar angle of the ground target are both $0$. 

\par
The MEO satellite constellation comprises $N_o$ orbits, with $N_M$ satellites on each orbit. The distance between the MEO satellite and Earth's center is denoted as $R_M$. The position of an orbit can be uniquely determined by its azimuth angle $\theta_{\parallel}$ and inclination angle $\theta_{\perp}$ relative to the ground target. Note that due to our focus on the position of the orbit relative to the ground target, these two angles differ from their definitions in astrodynamics. As shown in Fig.~\ref{figure1}, the inclination angle $\theta_{\perp}$ of an orbit is the angle between the positive $z$-axis direction and the normal vector of the orbital plane. The positive $z$-axis direction refers to the direction from the origin to the target. The probability density function (PDF) of $\theta_{\perp}$ is given as
\begin{equation}
    f_{\theta_{\perp}}(\theta) = \frac{\sin \theta}{2}, \ 0 \leq \theta < \pi.
\end{equation}
Then, we connect the origin to the point on the orbit closest to the target and obtain a segment. We project this segment onto the $xy$-plane, and the angle formed by the projected segment and the positive $x$-axis is referred to as the azimuth angle $\theta_{\parallel}$. Furthermore, $\theta_{\parallel}$ is assumed to follow a uniform distribution, i.e., $\theta_{\parallel} \sim \mathcal{U}[0, 2\pi)$. By making a simple extension of the conclusions from \cite{choi2024novel}, it can be proved that $\theta_{\perp}$ and $\theta_{\parallel}$ are also rotation invariant. 
As such, for an arbitrary ground reference position, the PDFs of $\theta_{\perp}$ and $\theta_{\parallel}$ relative to that position remain consistent.

\par
As for the spatial distribution of MEO satellites, the $N_M$ satellites on each orbit are uniformly distributed along the orbit, forming a one-dimensional BPP. According to \cite{choi2024novel}, the spatial distribution of orbits is mapped from a two-dimensional BPP. Therefore, the positions of MEO satellites in a single orbit are mapped from a one-dimensional BPP generated based on a two-dimensional BPP as the underlying point process. As a result, the positions of MEO satellites in the constellation follow a DSBPP.

\subsection{Antenna Pattern}\label{section2-3}
Given that satellites and the target are equipped with directional antennas. Satellites are equipped with wide spherical beams. The beam is directed toward the sub-satellite point, which is the intersection point between the Earth's surface and the line connecting the Earth's center to the satellite. Given the challenge of effectively covering targets situated at considerable distances from the sub-satellite point, we have established maximum central communication angles for LEO and MEO satellites, denoted by $\varphi_L$ and $\varphi_M$ respectively. As shown in Fig.~\ref{figure1}, when the angle formed between the line connecting the satellite to the user and the line connecting the satellite to the Earth's center exceeds $\varphi_L / 2$ or $\varphi_M / 2$, the satellite cannot be effectively used for precise positioning. In addition, the antenna gains of LEO satellites and MEO satellites are denoted as $G_L^t$ and $G_M^t$, respectively. 

\par
The ground target aligns the beams toward the associated satellites, including MEO and LEO satellites involved in positioning. The satellite association strategy, also known as the beam alignment strategy, will be demonstrated in Sec.~\ref{section4-2} Proposition~\ref{prop1}. For each beam, or for each alignment, the signal from the aligned satellite is the target signal and signals from all other satellites are considered interference. It is worth emphasizing that a satellite involved in positioning belongs to the target satellite for a specific beam or alignment, while at other times, it acts as an interfering satellite. 

\par
Then, we assume that the antenna patterns of the target are circularly symmetric, with the gain of a single antenna denoted as $G^r(\varphi^r)$. The aligned satellite receives the maximum beam gain $G_Q^m$, i.e., $G^r(0) = G_Q^m$. $\varphi^r$ is the dome angle between the associated satellite and the interfering satellite. Fig.~\ref{figure1} illustrates an example of $\varphi^r$. When the ground target aligns with a LEO satellite, $\varphi^r$ is formed between the line connecting the target to the associated LEO satellite and the line connecting the target to the interfering MEO satellite. The receiving beam gain $G^r(\varphi^r)$ decreases monotonically as $\varphi^r$ increases and the interference can be eliminated by directing the receiving beams toward the expected signals. For interference mitigation purposes, the directional receiving beam of the target shall be narrower than the satellite's wide beam.

\subsection{Channel Model}
We consider that the satellite-terrestrial link follows the free space fading model, experiencing large-scale fading and small-scale fading. This channel model framework is considered to be applicable to both LEO satellites and MEO satellites for downlinks to the ground \cite{choi2024novel}. In this case, the target's received power is \cite{talgat2020stochastic} 
\begin{equation} \label{rhor}
    \rho_Q^r = \left\{
    \begin{array}{ll}
    \rho_Q^t \, G_Q^t \, G^r(\varphi^r) \, \zeta \left( \frac{\lambda_Q}{4\pi} \right)^2 l^{-2} W, & l \leq d_Q^{\max}, \\
    \ \ \ \ \ \ \ \ \ \ \ \ \ \ 0, & \mathrm{otherwise},
	\end{array}
	\right.
\end{equation}
where $\rho_Q^t$, $\zeta$, $\lambda_Q$, and $l$ denote the transmit power, the system loss caused by the atmosphere, the wavelength, and the distance between satellite and target, respectively. $d_Q^{\max}$ is the maximum distance at which the target can detect the signals from satellites. Details of $d_Q^{\max}$ will be discussed in the next section. If (\ref{rhor}) represents the LEO satellite-target link, $Q$ is replaced by $L$; otherwise, for the MEO satellite-target link, $Q$ is replaced by $M$.

\par
Finally, we assume that the small-scale fading $W$ follows shadowed Rician (SR) fading, which is considered to be the most accurate small-scale fading model for space-terrestrial link \cite{abdi2003new}. The cumulative distribution function (CDF) of small-scale fading is given as follows:
\begin{equation}\label{SRCDF}
\begin{split}
    & F_W (w) = \left( \frac{2 b_0 m}{2 b_0 m + \Omega} \right)^m \sum_{z=0}^{\infty} \frac{(m)_z}{z! \Gamma(z+1)} \left( \frac{\Omega}{2 b_0 m + \Omega} \right)^z \Gamma_l \left(z+1, \frac{w}{2b_0} \right),
\end{split}
\end{equation}
where $(m)_z$ is the Pochhammer symbol, $m$, $b_0$, and $\Omega$ respectively denote the Nakagami parameter, average power of the line-of-sight (LoS) component, and the average power of the multi-path component. $\Gamma( \cdot )$ and $\Gamma_l(\cdot, \cdot )$ denote the gamma function and lower incomplete gamma function, respectively. Furthermore, the PDF of small-scale fading can be expressed as:
\begin{equation}
\begin{split}
    f_W(w) & = \left( \frac{2 b_0 m}{2 b_0 m + \Omega} \right)^m \frac{\exp\left( - \frac{w^2}{2b_0} \right)}{2b_0} {_1F_1} \left(m,1,\frac{\Omega m}{2b_0 (2b_0 m + \Omega)}\right),
\end{split}
\end{equation}
where ${_1F_1} \left( \cdot,\cdot,\cdot \right)$ is the Confluent Hypergeometric function.

\section{Localizability Analysis}
This section presents the definition and analytical expression of availability probability, contact angle distribution, interference power, and localizability probability. The theoretical framework is applicable not only to the analysis of hybrid LEO/MEO satellite constellations but also to pure LEO constellations or MEO constellations.

\subsection{Availability Probability}\label{section3-1}
This section derives the availability probability for hybrid LEO/MEO satellite networks. For ease of derivation, we first define the central angle.

\begin{definition}\label{def1}
    The central angle between A and B refers to the angle created by the line connecting A to the Earth's center and the line connecting B to the Earth's center.
\end{definition}

\par
To detect the signal from the satellite, the target needs to be within the satellite's main lobe. Furthermore, the target and the satellite need to establish a LoS link, i.e., the link cannot be blocked by the Earth. 
As such, we can derive the upper limit of central angle $\theta_Q^{\max}$ between the satellite and target, given in the following lemma.

\begin{lemma}\label{lemma1}
The maximum central angle at which the target can detect the signals from the satellite is 
\begin{align}\label{thetamax}
\begin{split}
    & \theta_Q^{\max} = \left\{
 	\begin{array}{ll}
    & \! \! \! \! \! \! \! \!  \arccos \left( \frac{R_{\oplus}}{R_Q} \right), \, {\mathrm{when}} \, \varphi_Q \geq 2 \arcsin \left( \frac{R_{\oplus}}{R_Q} \right), \\
    & \! \! \! \! \! \! \! \!  \arcsin\left( \frac{R_Q \sin\left( {\varphi_Q}/{2} \right)}{R_{\oplus}} \right) - \frac{\varphi_Q}{2}, \ \ \ \, {\mathrm{otherwise}}.
	\end{array}
	\right.
\end{split}
\end{align}
where $Q$ is substituted with $L$ for LEO satellites and $M$ for MEO satellites.
\end{lemma}
\begin{IEEEproof}
    See Appendix~\ref{app:lemma1}.
\end{IEEEproof}

\par
Then, we can establish the relationship between the maximum detectable central angle $\theta_Q^{\max}$ and the maximum detectable distance $d_Q^{\max}$, as 
\begin{equation}
    d_Q^{\max} = \sqrt{ R_Q^2 + R_{\oplus}^2 - 2 R_Q R_{\oplus} \cos\theta_Q^{\max} }, \, Q=\{L,M\}.
\end{equation}

Consequently, we define the following $K-$availability probability.

\begin{definition} [$K-$Availability Probability] \label{definition3}
    $K-$availability probability refers to the probability of having at least $K$ satellites available for the ground target. When an LEO or MEO satellite is available, it has a distance less than $d_L^{\max}$ or $d_M^{\max}$ from the target.
\end{definition}

As shown in Definition~\ref{definition3}, $d_Q^{\max}$ can make the definition more intuitively and clearly expressed. However, as will be clear from the upcoming analysis, using central angles $\theta_Q^{\max}$ instead of Euclidean distance $d_Q^{\max}$ is more concise in derivations.

\begin{lemma}\label{lemma2}
$K-$availability probability for LEO satellites can be calculated as,
\begin{equation}\label{availLEO}
\begin{split}
& P_L^A (K) = 1 - \sum_{k=0}^{K-1} \binom{N_L}{k} \left( \frac{1 - \cos\theta_L^{\max}}{2} \right)^k \left( \frac{1 + \cos\theta_L^{\max}}{2} \right)^{N_L-k}.
\end{split}
\end{equation}
\end{lemma}
\begin{IEEEproof}
See Appendix~\ref{app:lemma2}.
\end{IEEEproof}

After computing the $K-$availability probability for LEO satellites, the next step is to derive the same quantity for MEO satellites. First, we present the maximum central angle of a given orbit in Lemma~\ref{lemma3}.

\begin{lemma}\label{lemma3}
Given that two satellites are located in an orbit with inclination angle $\theta_\perp$, and these two satellites are at a distance less than $d_M^{\max}$ from the target, the maximum central angle between them is
\begin{align}\label{thetacmax}
\begin{split}
    & \theta_{c,M}^{\max} \left( \theta_{\perp},d_M^{\max} \right)  = \left\{
 	\begin{array}{ll}
    &  \! \! \! \! \! \!  2 \arccos \left( \frac{R_{\oplus}^2 + R_M^2 - (d_M^{\max})^2}{2R_{\oplus} R_M \sin\theta_{\perp}} \right), \\
    &  {\mathrm{when}} \left| \theta_\perp - \frac{\pi}{2} \right| \leq \arccos \left( \frac{R_\oplus^2 + R_M^2 - (d_M^{\max})^2}{2 R_\oplus R_M} \right), \\
    & \! \!  0, \ \ {\mathrm{otherwise}}.
	\end{array}
	\right.
\end{split}
\end{align}
\end{lemma}
\begin{IEEEproof}
    See Appendix~\ref{app:lemma3}.
\end{IEEEproof}

Lemma~\ref{lemma1} can be regarded as the unconditional upper limit of the central angle, while Lemma~\ref{lemma3} represents the upper limit of the central angle given a specific orbit. Based on Lemma~\ref{lemma3}, the $K-$availability of MEO satellites is derived as follows.

\begin{lemma}\label{lemma4}
$K-$availability probability for MEO satellites is
\begin{equation}\label{availMEO}
P_M^A (K) = 1 - \sum_{k=0}^{K-1} \binom{N_M N_O}{k} \left( p_{M,1}^A \right)^k \left( 1 - p_{M,1}^A \right)^{N_M N_O-k},
\end{equation}
where $p_{M,1}^A$ is the probability that a given MEO satellite is available for the target, given by
\begin{equation}
    p_{M,1}^A = \int_{\frac{\pi}{2} - \arccos \left( \frac{R_\oplus^2 + R_M^2 - (d_M^{\max})^2}{2 R_\oplus R_M} \right)}^{ \frac{\pi}{2} + \arccos \left( \frac{R_\oplus^2 + R_M^2 - (d_M^{\max})^2}{2 R_\oplus R_M} \right) } \frac{ \theta_{c,M}^{\max} \left( \theta,d_M^{\max} \right) \sin\theta}{4 \pi} {\mathrm{d}} \theta.
\end{equation}
\end{lemma}
\begin{IEEEproof}
See Appendix~\ref{app:lemma4}.
\end{IEEEproof}

Finally, the following theorem provides the analytical expression of $K-$availability probability based on the Lemma~\ref{lemma2} and Lemma~\ref{lemma4}.

\begin{theorem}\label{theorem1}
The $K-$availability probability for hybrid LEO/MEO satellite networks is given by
\begin{equation} \label{PallA}
P_{\mathrm{all}}^A(K) = \! \! \! \! \! \! \!  \! \! \! \! \! \! \! \! \sum_{k=0}^{\min\{K-1,N_M^{\max}-1\}} \! \! \! \!  \! \! \! \! \! \! \! \! \! P_L^A ( K -k) \, p_M^A (k) + \! \! \! \! \! \! \! \sum_{k=K}^{\max\{K,N_M^{\max}\}} \! \! \! \! p_M^A (k),
\end{equation}
where $N_M^{\max}$ is predefined upper limit of available MEO satellites, and $P_L^A (K)$ is given in Lemma~\ref{lemma2}. $p_M^A (k)$ is the probability that exactly $k$ MEO satellites are available for the target,
\begin{equation}
p_M^A (k) = \binom{N_M N_O}{k} \left( p_{M,1}^A \right)^k \left( 1 - p_{M,1}^A \right)^{N_M N_O-k}.
\end{equation}
\end{theorem}
\begin{IEEEproof}
See Appendix~\ref{app:theorem1}.
\end{IEEEproof}

$N_M^{\max} \ll N_M N_O$ is set to reduce computational complexity. 
The satellite deployment density in MEO constellations is usually lower than that in LEO constellations. 
Therefore, setting an upper limit for MEO satellites is more effective in reducing computational complexity compared to LEO satellites. 

\begin{proposition}\label{prop2}
A criterion regarding the recommended value for $N_M^{\max}$ is given as follows:
\begin{equation}
    N_M^{\max} = \underset{K}{\argmin} \left\{ 1 - \sum_{k=0}^K p_M^A(k) \leq \varepsilon \right\},
\end{equation}
where $\varepsilon$ is a positive real value close to $0$. 
\end{proposition}

We set $\varepsilon = 0.01$ in this article; namely, the probability of having more than $N_M^{\max}$ available MEO satellites is less than $1\%$.

\subsection{Contact Angle Distribution} \label{section4-2}
This subsection elaborates on the association strategy for the ground target and the contact angle distributions for LEO and MEO satellites. In satellite-enabled positioning systems, accuracy can be enhanced by utilizing consecutive positioning results from the same satellite, e.g., the Kalman filtering technique employed in the GPS system \cite{almagbile2010evaluating}. To minimize the cost of beam scanning, a beam is generally associated with a specific satellite during a period rather than switching satellites frequently. In the following proposition, we provide the association strategy adopted by this article.

\begin{proposition}\label{prop1}
    Considering the maturity of the positioning technology of MEO satellites, the beams of the ground target are prioritized to associate with available MEO satellites, while the remaining beams are sequentially associated with the nearest LEO satellites.
\end{proposition}

This association strategy is designed from the perspective of an LEO-assisted MEO satellite-enabled positioning system. Notice that the localizability analysis based on the associated strategy depends on the relative distance between the associated satellite and the target. Therefore, we propose the concept of the contact angle to measure this relative distance.

\begin{definition}[Contact Angle]\label{definition4}
    The central angle between the target and its $k^{th}$ nearest LEO satellite is referred to as the $k$-LEO contact angle. The central angle between an MEO satellite and the target is referred to as the MEO contact angle.
\end{definition}

Then, the PDFs of the contact angle distributions are derived in the following lemmas.

\begin{lemma}\label{lemma5}
The PDF of the $K-$LEO contact angle distribution is given as, 
\begin{equation}
\begin{split}
    & f_{\theta_L}^{(K)} (\theta) = 1 - \frac{\sin\theta}{2^{N_L}} \sum_{n=0}^{K-1} \binom{N_L}{n} \left( 1- \cos \theta \right)^n \\  
    & \times \left( 1+\cos\theta \right)^{N_L-n} \left( \frac{n}{1-\cos\theta} - \frac{N_L - n}{1+\cos\theta} \right),
\end{split}
\end{equation}
where $0 \leq \theta \leq \theta_L^{\max}$.
\end{lemma}
\begin{IEEEproof}
See Appendix~\ref{app:lemma5}.
\end{IEEEproof}

\begin{lemma}\label{lemma6}
The PDF of the MEO contact angle distribution is given as, 
\begin{equation}
\begin{split}
    & f_{\theta_M} (\theta) =  \frac{\sin\theta}{\pi} \int_{\frac{\pi}{2}-\theta}^{\frac{\pi}{2}} \frac{1}{\sqrt{\sin^2\theta_{\perp} - \cos^2\theta}} {\mathrm{d}}\theta_{\perp},
\end{split}
\end{equation}
where $0 \leq \theta \leq \theta_M^{\max}$.
\end{lemma}
\begin{IEEEproof}
See Appendix~\ref{app:lemma6}.
\end{IEEEproof}

Toward this end, we have finished the analysis from the topological perspective. 
In the next subsection, we delve into the channel aspect and introduce $K-$localizability probability for further analysis.

\subsection{Localizability Probability}
This subsection provides the definition of $K$-localizability probability and derives the analytical expression of interference power and $K$-localizability probability.

\begin{definition}[$K-$Localizability Probability]\label{definition5}
    $K-$localizability probability is defined as the probability that the $K$ associated satellites' instantaneous received Signal-to-interference-plus-noise ratios (SINRs) at the target are all greater than the threshold $\gamma_L$ for LEO satellites or $\gamma_M$ for MEO satellites. 
\end{definition}

The mathematical definition of $K-$localizability probability can be expressed by the following formula:
\begin{equation}
    P_{\mathrm{all}}^L (K) = \prod_{k=1}^K \mathbbm{P}\left[ \mathrm{SINR}_k = \frac{\rho_{Q,k}^r}{I_k +\sigma_Q^2} > \gamma_Q \right],
\end{equation}
where $\mathrm{SINR}_k$ is the instantaneous received SINR of the $k^{th}$ associated satellite. $Q$ is replaced with $M$ when the $k^{th}$ associated satellite is an MEO satellite; otherwise, $Q$ is replaced with $L$. $\rho_{Q,k}^r$ is the received power of the $k^{th}$ associated satellite. Its randomness is caused by the contact angle distribution and SR fading. $\sigma_Q^2$ is a constant representing the noise power. Moreover, $I_k$ is the total interference power received by the target when the beam is aligned with the $k^{th}$ associated satellite. Apart from the aligned satellite, the received power from other satellites is regarded as interference. 
Notably, since the LEO and MEO satellite-enabled positioning systems use different carrier frequencies, there is no mutual co-channel interference \cite{ferre2021comparison}. The following lemma estimates the approximate distribution of interference power.

\begin{lemma}\label{lemma7}
Given that the central angle between the associated satellite and the target is $\theta$, the interference power of the LEO satellite constellation can be expressed as a mixed random variable $I(\theta)$. The probability that $I(\theta)$ can be approximated as $0$ is 
\begin{equation} 
    \mathbbm{P}[I(\theta) \approx 0] = \left( \frac{1 + \cos\theta_d^{\max}}{2} \right)^{N_L},
\end{equation}
where $\theta_d^{\max}$ can be expressed as,
\begin{equation}\label{thetadmax}
\begin{split}
    & \theta_d^{\max} = \cot^{-1} \Bigg( \frac{R_L^2}{R_L^2 - R_{\oplus}^2} \bigg( \cot \varphi_{\max} \\
    & + 
    \sqrt{(R_{\oplus}/R_L)^2 \left(1 + \cot^2\varphi_{\max} - (R_{\oplus}/R_L)^2 \right) } \bigg) \Bigg). 
\end{split}
\end{equation}
$\varphi_{\max}$ in (\ref{thetadmax}) is defined as the effective receiving beam range.

\par
When $I(\theta) \neq 0$, the PDF of $I(\theta)$ is shown in (\ref{PDFI}) at the top of the next page. 

\begin{table*}
\begin{equation}\label{PDFI}
\begin{split}
    & f_{I} (I) = \int_{0}^{\theta_d^{\max}}  f_W \left(  \frac{I \left(R_L^2 + R_{\oplus}^2 - 2 R_L R_{\oplus} \cos\theta_L \right) }{\rho_L^t G_L^t  \zeta \left( \frac{\lambda_L}{4\pi}\right)^2 G^r \left( \cot^{-1} \left( \cot\theta_I - \frac{R_{\oplus}}{R_L} \sqrt{1+\cot^2\theta_I} \right) \right)} \right) \\
    & \times \frac{\left(R_L^2 + R_{\oplus}^2 - 2 R_L R_{\oplus} \cos\theta_L \right) }{\rho_L^t G_L^t  \zeta \left( \frac{\lambda_L}{4\pi}\right)^2 G^r \left( \cot^{-1} \left( \cot\theta_I - \frac{R_{\oplus}}{R_L} \sqrt{1+\cot^2\theta_I} \right) \right)} \, \frac{\sin\theta_I}{1-\cos \theta_d^{\max}} \, {\mathrm{d}}\theta_I.
\end{split}
\end{equation}
\hrule
\end{table*}
\end{lemma}
\begin{IEEEproof}
See Appendix~\ref{app:lemma7}.
\end{IEEEproof}

It is noteworthy that the interference power cannot actually be zero. 
However, under the premise of directional reception, when the interference power is negligible compared to noise, we can approximate it as zero. 
Due to the much lower density of MEO satellite deployment compared to LEO, as indicated in Appendix~\ref{app:lemma7}, the probability of interference power being approximated as $0$ is high. Furthermore, it is easier to avoid co-channel interference through frequency coordination. Therefore, we neglect the impact of interference from MEO constellations on the localizability. Based on the contact angle distributions and interference power distribution, the analytical expressions of $K$-localizability probability can be given in the following lemmas and theorem.

\begin{lemma}\label{lemma8}
$K-$localizability probability for LEO satellites can be calculated as,
\begin{sequation}
\begin{split}
    & P_L^C(K) = \prod_{k=1}^K \int_0^{\theta_L^{\max}} f_{\theta_L}^{(k)}(\theta) \int_0^{\infty} f_I(I) \Bigg( 1 - F_W \Bigg( \left(  \frac{4\pi}{\lambda_L} \right)^2 \\
    & \times \frac{\gamma_L \left(I+\sigma_L^2\right) \left( R_L^2 + R_{\oplus}^2 - 2R_L R_{\oplus} \cos\theta_L \right) }{\rho_L^t G_L^t G_L^m \zeta} \Bigg) \Bigg) {\mathrm{d}}I {\mathrm{d}}\theta_L.
\end{split}    
\end{sequation}
\end{lemma}
\begin{IEEEproof}
See Appendix~\ref{app:lemma8}.
\end{IEEEproof}

The analytical expression in the above lemma can be regarded as an estimation of $K$-localizability for the hybrid LEO/MEO satellite network when no MEO satellites are available. On the contrary, the following lemma corresponds to the $K$-localizability when localization can be achieved without the assistance of LEO satellites.

\begin{lemma}\label{lemma9}
$K-$localizability probability for MEO satellites can be calculated as, 
\begin{equation}
P_M^C(K) = 1 - \sum_{k=0}^{K-1} \binom{N_M N_O}{k} \left( p_{M,1}^C \right)^k \left( 1 - p_{M,1}^C \right)^{N_M N_O-k},  
\end{equation}
where $p_{M,1}^C$ is the probability that the target is  localizable for a given MEO satellite,
\begin{equation}
\begin{split}
    & p_{M,1}^C = \int_0^{\theta_M^{\max}} f_{\theta_M}(\theta) \Bigg( 1 - F_W \Bigg( \gamma_M \left(  \frac{4\pi}{\lambda_M} \right)^2 \\
    & \times \frac{ \sigma_M^2 \left( R_M^2 + R_{\oplus}^2 - 2R_M R_{\oplus} \cos\theta_M \right) }{\rho_M^t G_M^t G_M^m \zeta} \Bigg) \Bigg) {\mathrm{d}} \theta_M.
\end{split}    
\end{equation}
\end{lemma}
\begin{IEEEproof}
The proof of Lemma~\ref{lemma9} is similar to that of Lemma~\ref{lemma4} and Lemma~\ref{lemma8}, therefore omitted here.
\end{IEEEproof}

Next, we provide the results of the localizability probability for hybrid satellite networks.

\begin{theorem}\label{theorem2}
$K-$localizability probability for hybrid LEO/MEO satellite networks can be calculated as,
\begin{equation}
P_{\mathrm{all}}^C (K) = \! \! \! \! \! \! \!  \! \! \! \! \! \! \! \! \sum_{k=0}^{\min\{K-1,N_M^{\max}-1\}} \! \! \! \!  \! \! \! \! \! \! \! \! \! P_L^C ( K -k) \, p_M^C (k) + \! \! \! \! \! \! \! \sum_{k=K}^{\max\{K,N_M^{\max}\}} \! \! \! \! p_M^C (k),
\end{equation}
where $N_M^{\max}$ is defined in Theorem~\ref{theorem1} and Proposition~\ref{prop2}. $p_M^C (k)$ is the probability that the target is localizable for exactly $k$ MEO satellites,
\begin{equation}
p_M^C (k) = \binom{N_M N_O}{k} \left( p_{M,1}^C \right)^k \left( 1 - p_{M,1}^C \right)^{N_M N_O-k}.
\end{equation}
\end{theorem}
\begin{IEEEproof}
The proof of Theorem~\ref{theorem2} is similar to that of Theorem~\ref{theorem1}, therefore omitted here.
\end{IEEEproof}

The above results demonstrate that the expressions for the localizability probability, whether in LEO, MEO, or hybrid constellations, all contain only double integrals, indicating an acceptable level of computational complexity. Finally, as shown in the Theorem~\ref{theorem1} and the Theorem~\ref{theorem2}, the $K$-availability probability and $K$-localizability probability of the hybrid LEO/MEO satellite network involve single integral and double integrals, respectively. Considering the complex distance analysis due to spherical topology and the complex interference analysis caused by beam pointing, the analytical results in this paper are computationally inexpensive, with a complexity equivalent to or lower than that of most current research in the domain of spherical SG.


\begin{table*}[ht]
\centering
\caption{System parameters \cite{talgat2020stochastic,ferre2022leo,singhal2023leo}.}
\label{table2}
\renewcommand{\arraystretch}{1.1}
\begin{tabular}{|c|c|c|}
\hline
Notation   &  Physical meaning  & Default value    \\ \hline \hline
$h_L$, $h_M$   & Altitude of the satellite relative to the Earth's surface  & $1000$~km, $20000$~km    \\ \hline
$N_L$   & Number of LEO satellites   & $2000$  \\ \hline
$N_o$  & Number of orbits (for MEO satellites)  & $2$  \\ \hline
$N_M$   & Number of satellites on an orbit (for MEO satellites) & $6$  \\ \hline
$\varphi_L$, $\varphi_M$   & Maximum central communication angle  & $\pi/4$, $\pi/6$  \\ \hline
$\lambda_L$, $\lambda_M$   & Wavelength & $0.0150$~m, $0.190$~m  \\ \hline
$\rho_L^t$, $\rho_M^t$   & Transmission power & $15$~dBW, $18$~dBW    \\ \hline
$G_L^t$, $G_M^t$   & Transmitting antenna gain & $33.8$~dBi, $24.1$~dBi  \\ \hline
$G_L^m$, $G_M^m$ & Maximum receiving antenna gain  &  $31.8$~dBi, $5$~dBi  \\ \hline
$\zeta$   & System loss   & $-6$~dB     \\ \hline
$\sigma_L^2, \sigma_M^2$     & Noise power   & $-90.2$~dBm, $-103.9$~dBm  \\  \hline
$\gamma_L$, $\gamma_M$ &  Coverage threshold & $10$~dBi, $-16$~dBi \\ \hline
$\varphi_{\mathrm{3dB}}$ &  Half-power beamwidth & $8$~degree \\ \hline
$ (\Omega,b_0,m)$ & Parameters of SR fading   & $ (1.29,0.158,19.4)$ \\ \hline
\end{tabular}
\end{table*}

\section{Numerical Results}
This section presents the numerical results of $K$-availability probability and $K$-localizability probability. The overlap between the results obtained from Monte Carlo simulations (lines) and the derived analytical results (marks) in the figures validates the accuracy of the analytical results in this article.

\subsection{Parameter Setting}
To visually demonstrate the impact of altitude on performance, in this section, we define $h_Q = R_Q - R_\oplus$ as the altitude of the satellite relative to the Earth's surface, where $Q = \{L, M\}$. In addition, we apply the Gaussian antenna pattern \cite{gagliardi2012satellite} for the target's beam, and the receiving gain can be expressed as
\begin{equation}
    G^r(\varphi^r) = G_Q^m \, 2^{-(\varphi^r)^2/ \varphi_{\rm 3dB}^2},
\end{equation}
where $\varphi_{\mathrm{3dB}}$ denotes half-power beamwidth of the receiving beam. Unless otherwise stated, other parameters will be set to their default values as Table~\ref{table2}. The following are some explanations of the setting in Table~\ref{table2}. The configuration of the number of orbits $N_o$ and the number of satellites on an orbit $N_M$ applies only to MEO satellite constellations. 

\par
Considering that satellite systems typically focus more on frequency rather than wavelength, we declare that in the numerical results corresponding to the cases in this section, the LEO satellites are concerned with the $20$~GHz frequency band, i.e., the Ka-band, while the MEO satellites operate at $1.579$~GHz, i.e., the L-band. In this case, the corresponding Wavelengths in the above analytical framework are set as $\lambda_L = 0.0150$~m and $\lambda_M= 0.190$~m.

\subsection{Availability Probability}
\begin{figure}[ht]
	\centering
    \includegraphics[width = 0.6\linewidth]{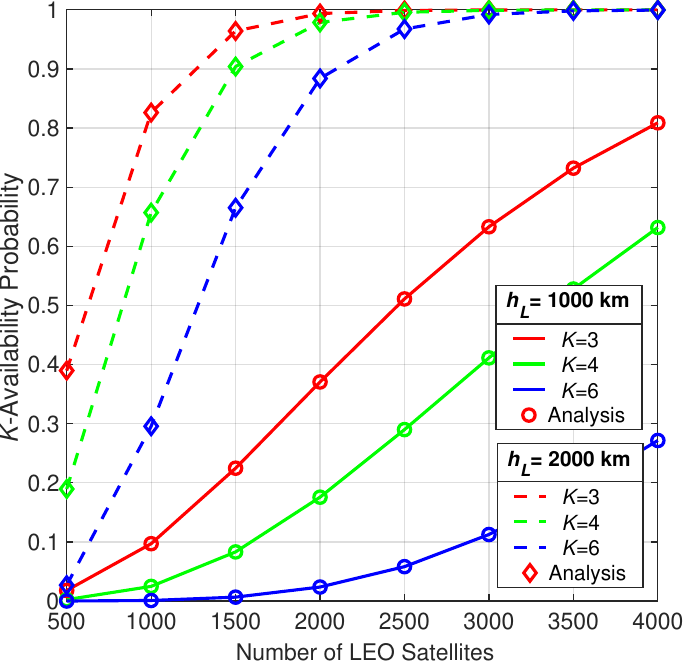}
	\caption{$K-$availability probability for LEO satellites.}
	\label{figure2}
\end{figure}

This subsection provides the numerical results of $K$-availability probabilities for LEO, MEO, and hybrid satellite networks. As shown in Fig.~\ref{figure2}, the altitude of the satellite constellation $h_s$ has a significant impact on the $K$-availability probability for LEO satellites. When $h_s=2000$~km, a constellation of more than $3000$ LEO satellites can provide at least $6$ available satellites. However, even with $4000$ satellites, a mega constellation with altitude $h_s=1000$~km still cannot completely ensure the basic positioning requirements (typically, it requires $3$ or $4$ available satellites for positioning, depending on the specific positioning technology).

\par
\begin{figure}[ht]
	\centering
    \includegraphics[width = 0.6\linewidth]{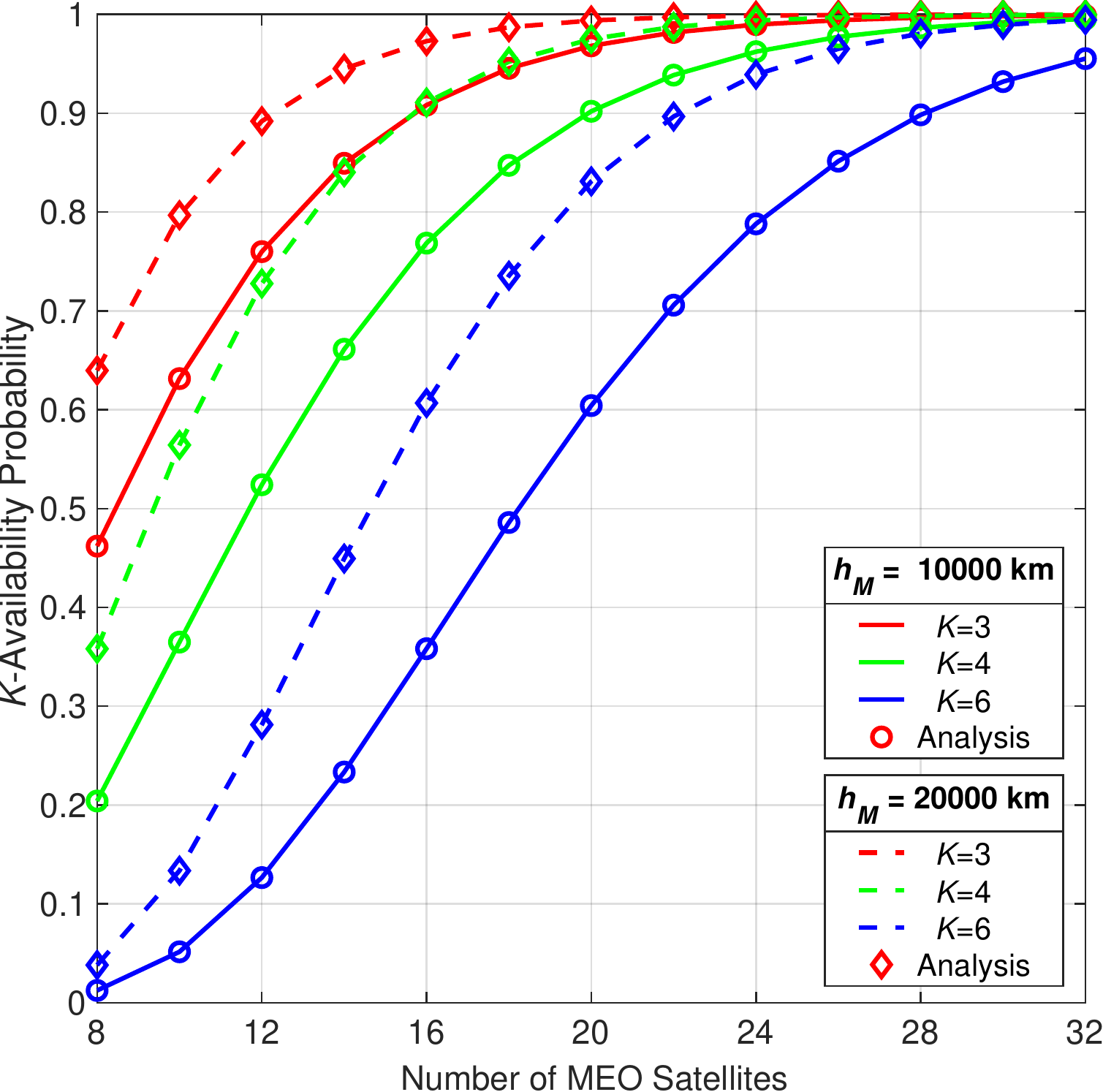}
	\caption{$K-$availability probability for MEO satellites.}
	\label{figure3}
\end{figure}

As revealed by Fig.~\ref{figure3}, the number of MEO satellites required to achieve the same availability probability is much lower than that of LEO satellites. At an altitude of $20000$~km, $24$ satellites (the minimum operational number for GPS) can guarantee $4-$availability, while $31$ satellites (the typical operational number for GPS) can generally meet $6-$availability. With the same number of satellites, deploying them at $20000$~km altitude results in a $5\%$ to $15\%$ increase in availability probability compared to $10000$~km altitude.

\par
\begin{figure}[ht]
	\centering
    \includegraphics[width = 0.6\linewidth]{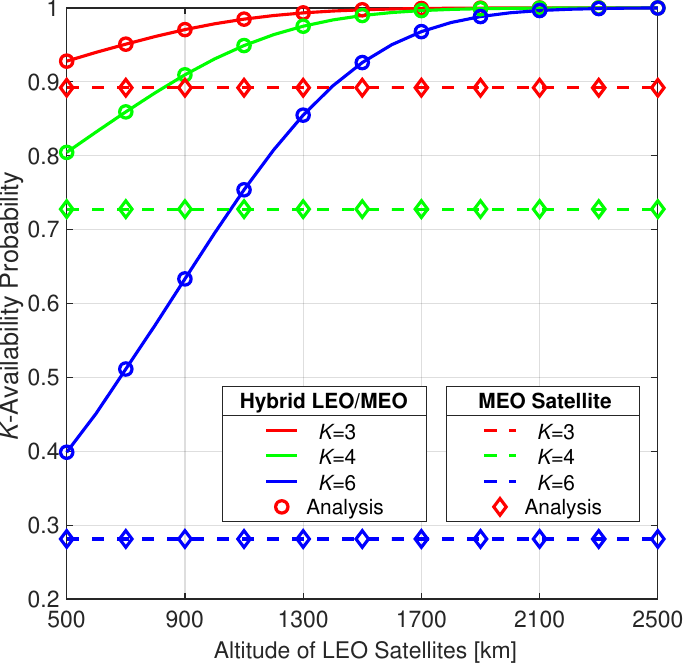}
	\caption{$K-$availability probability for hybrid LEO/MEO satellite networks.}
	\label{figure4}
\end{figure}

Fig.~\ref{figure4} examines the impact of altitude on a hybrid LEO/MEO satellite constellation. Compared to a constellation with $12$ MEO satellites at an altitude of $h_M = 20000$~km, the hybrid constellation with extra $2000$ LEO satellites at an altitude of $h_L = 1500$~km shows a $10\%,28\%,64\%$ increase in $3,4,6-$availability probability, respectively. We also find that when the availability probability is low, increasing the altitude of the LEO satellite constellation can bring a significant improvement to satellite availability.

\par
\begin{figure}[ht]
	\centering
    \includegraphics[width = 0.7\linewidth]{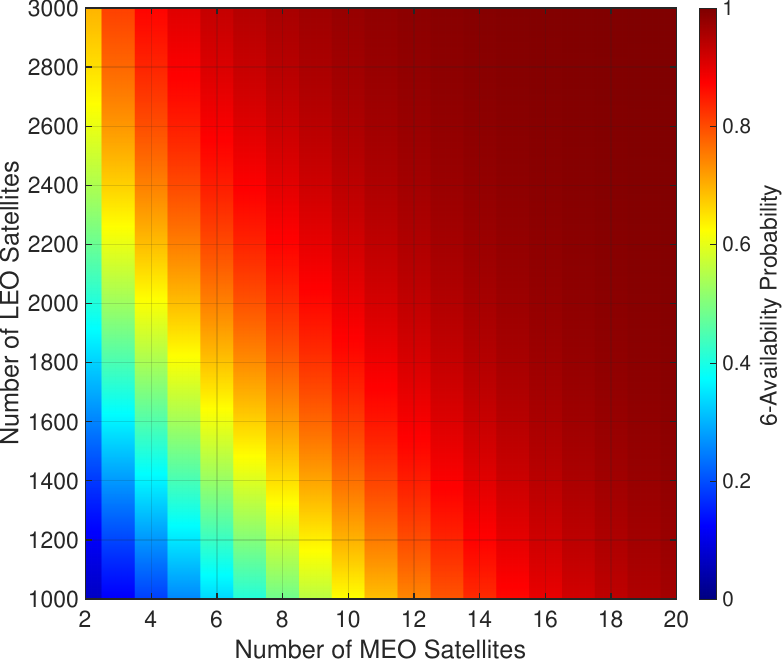}
	\caption{Heatmap of $6-$availability probability for hybrid LEO/MEO satellite networks.}
	\label{figure5}
\end{figure}

Fig.~\ref{figure5} displays the influence of the number of satellites on $6-$availability probability using a heatmap. 
The values of the points on the heatmap are calculated based on Theorem~\ref{theorem1} (whose accuracy has been validated). The color boundaries in Fig.~\ref{figure5} are almost parallel to the main diagonal of the heatmap, indicating that the MEO and LEO satellites have a complementary relationship in terms of achieving availability. Specifically, for each MEO satellite decreased, it is necessary to add around $150$ LEO satellites to maintain the $6-$availability probability for the hybrid constellation.

\subsection{Localizability Probability}
\begin{figure}[ht]
	\centering
    \includegraphics[width = 0.6\linewidth]{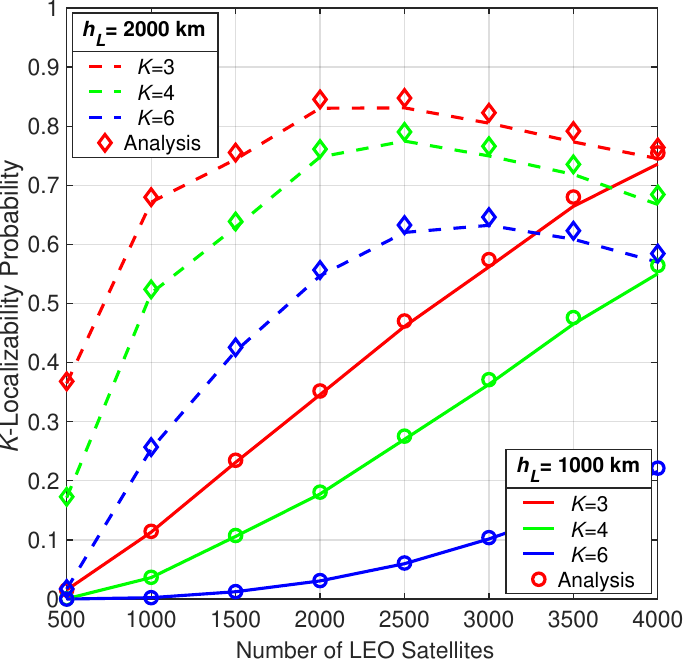}
	\caption{$K-$localizability probability for LEO satellites.}
	\label{figure6}
\end{figure}

This subsection provides the numerical results of $K$-localizability probabilities for LEO, MEO, and hybrid satellite networks. 
Fig.~\ref{figure6} reveals that with an increase in the number of LEO satellites, the $K-$localizability probability exhibits two different trends depending on the altitude of satellite deployment. 
Specifically, when $h_L = 1000$~km, availability remains the dominant factor in determining the $K-$localizability probability. Therefore, as the number of satellites increases, the $K-$localizability probability monotonically increases. When $h_L = 2000$~km, the $K-$localizability probability initially increases due to the rise in $K-$availability probability. However, once availability nears 100$\%$, interference becomes the decisive factor affecting the $K-$localizability probability, leading to a decline in the $K-$localizability probability as the number of satellites continues to increase. When the number of satellites is less than $4000$, deploying them at a relative altitude of $h_L=2000$~km has an advantage in terms of localizability compared to $h_L=1000$~km.

\par
\begin{figure}[ht]
	\centering
    \includegraphics[width = 0.6\linewidth]{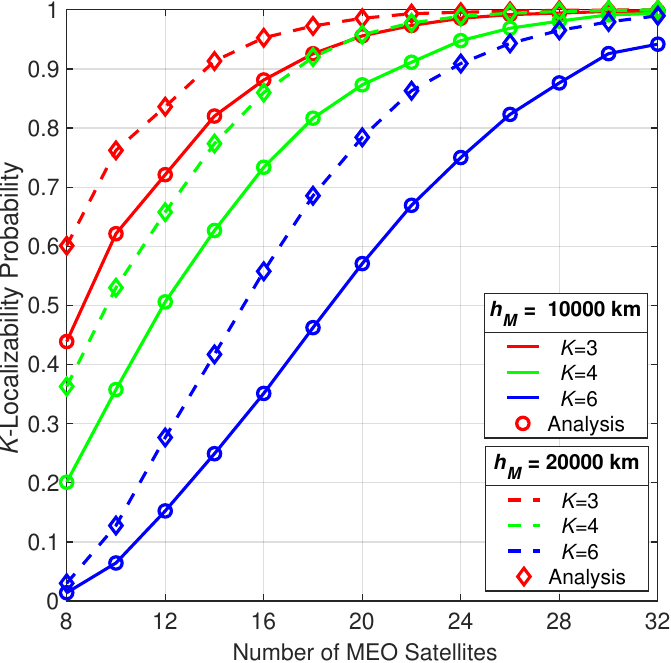}
	\caption{$K-$localizability probability for MEO satellites.}
	\label{figure7}
\end{figure}

Comparing the numerical results in Fig.~\ref{figure6} and Fig.~\ref{figure7}, we can conclude that a constellation with $30$ MEO satellites has an advantage over a constellation with $3000$ LEO satellites in terms of localizability performance. A constellation of more than $30$ MEO satellites is already sufficient to ensure that both the $4-$availability probability and the $4-$localizability probability are close to $1$. Additionally, an interesting finding is that changing the altitude of the MEO satellite constellation has little impact on localizability. The reason behind this phenomenon is that increasing the MEO constellation's altitude $h_M$, while enhancing availability, also weakens the received signal strength.

\par
\begin{figure}[ht]
	\centering
    \includegraphics[width = 0.6\linewidth]{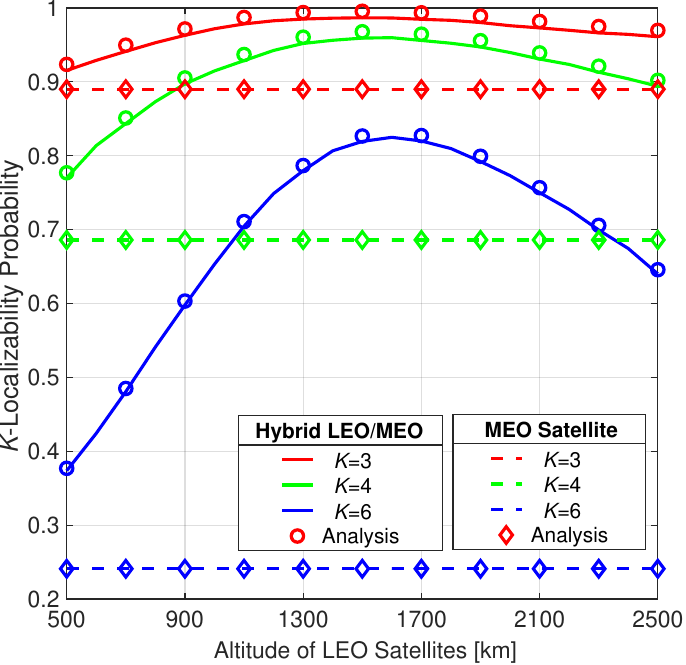}
	\caption{$K-$localizability probability for hybrid LEO/MEO satellite networks.}
	\label{figure8}
\end{figure}

In Fig.~\ref{figure8}, compared to a constellation with $12$ MEO satellites at an altitude of $h_M = 20000$~km, the hybrid constellation with extra $2000$ LEO satellites shows a gain in $3,4,6-$localizability probability up to $10\%,27\%,60\%$, respectively. Due to the balance of availability and received signal strength, deploying the LEO satellite constellation at an altitude of about $h_L=1500$~km can achieve the maximum gain in localizability performance. 

\par
\begin{figure}[ht]
	\centering
    \includegraphics[width = 0.7\linewidth]{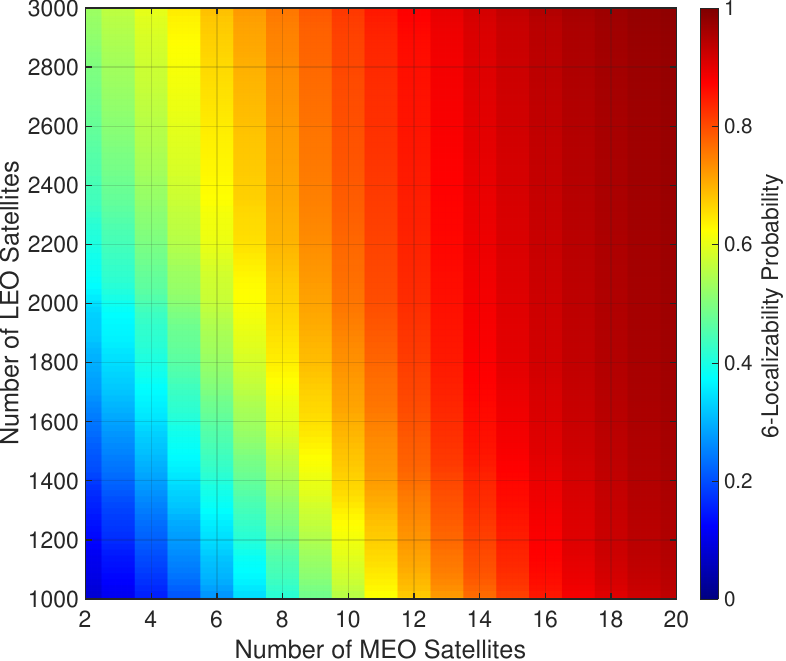}
	\caption{Heatmap of $6-$localizability probability for hybrid LEO/MEO satellite networks.}
	\label{figure9}
\end{figure}

Fig.~\ref{figure9} demonstrates the influence of numbers of LEO and MEO satellites on the $6-$localizability probability, where $h_L=2000$~km and $h_M=20000$~km. In terms of localizability, the two types of satellites still show a complementary relationship in terms of quantity but do not exhibit a typical linear relationship as that in availability.

\section{Further Discussion}
Building on the aforementioned statements, the primary contribution of this paper lies in proposing an analytical framework for studying satellite-enabled positioning issues and the DSBPP model applicable to MEO satellite constellations. However, there are limitations in various aspects such as detailed modeling of channels and beams, design of positioning algorithms-based accuracy evaluation, and research on models based on SG. Therefore, we have dedicated this section further to delve deeper into these issues, which will also serve as a guide for future research.

\subsection{Antenna Pattern}
This subsection first presents four types of basic antenna patterns with their beam gain models in Table~\ref{table1}, then designs a case study on the antenna patterns, followed by a discussion of the limitations of the beam-related analysis in this paper, serving as a direction for future improvements. 

\begin{table}[ht]
\centering
\caption{Beam gains of basic antenna patterns.}
\label{table1}
\renewcommand\arraystretch{1.1}
\resizebox{0.6\linewidth}{!}{
\begin{tabular}{|c|c|}
\hline
Antenna Pattern   & Model \\ \hline
Gaussian \cite{gagliardi2012satellite}    & $G^r(\varphi^r) = G_L^m \, 2^{-(\varphi^r)^2/ \varphi_{\rm 3dB}^2}$  \\ \hline
Flat-top \cite{balanis2015antenna} &
  $ G^r(\varphi^r) = \left\{\begin{matrix} G_L^m  & \left | \varphi^r \right | \leq  \varphi_{\rm 3dB} \\  0    &  { \rm otherwise} \end{matrix}\right.$ 
   \\ \hline
    Sinc \cite{yu2017coverage} & $ G^r(\varphi^r) = G_L^m \frac{\sin^2\left ( \pi N_a \varphi^r \right )}{\left ( \pi N_a \varphi^r \right )^2}$  \\ \hline
    Cosine \cite{yu2017coverage} &
    $ G(\varphi^r) = \left\{\begin{matrix} G_L^m   \cos^2\left ( \frac{\pi N_a}{2} \varphi^r \right )  &  \left | \varphi^r \right |   \leq  \frac{1}{N_a}\\ 0    &  {\rm otherwise} \end{matrix}\right.$
   \\ \hline
\end{tabular}
}
\end{table}

\par
In the numerical simulations of the previous section, the receiving beam gain was designed according to a Gaussian antenna pattern. Therefore, we extend this assumption to demonstrate that the analytical framework in this paper can be applied to various antenna patterns. As shown in Table~\ref{table1}, $N_a$ is defined as the number of antenna elements. Recall that $G_L^m$ and $\varphi_{\mathrm{3dB}}$ denote the maximum receiving antenna gain and the half-power beamwidth. 

\par
\begin{figure}[ht]
	\centering
    \includegraphics[width = 0.6\linewidth]{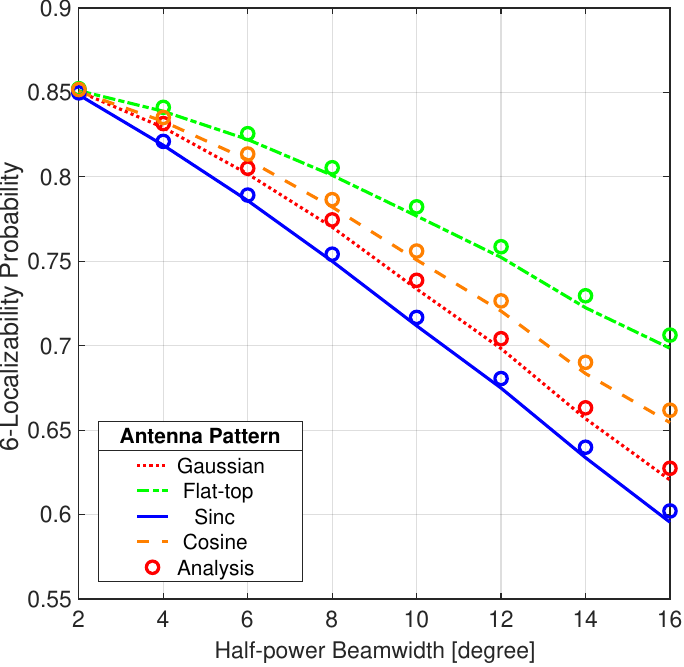}
	\caption{Influence of attenna pattern on $6-$localizability probability.}
	\label{figure10}
\end{figure}

Next, we compare the impact of different antenna patterns on localizability probability in Fig.~\ref{figure10}. To ensure consistency in the effective beamwidth, we set $N_a = 35$. $\varphi_{\max}$ follows the default values in Table~\ref{table2}, while $N_a$ is set to $35$. The effective receiving beam range $\varphi_{\max}$ defined in (\ref{thetadmax}) under different antenna patterns are given by
\begin{align}
\begin{split}
    \varphi_{\max} = \left\{
 	\begin{array}{ll}
    & \! \! \! \! \! \! \!  3 \varphi_{\mathrm{3dB}}, \ \ {\mathrm{Gaussian}}, \\
    & \! \! \! \! \! \varphi_{\mathrm{3dB}}, \ \ {\mathrm{Flat-top}}, \\
    & \! \! \!  \frac{3}{N_a}, \ \ \ \ \ \ \  {\mathrm{Sinc}}, \\
    & \! \! \!  \frac{1}{N_a}, \ \ \ \ \ \ {\mathrm{Cosine}}.
	\end{array}
	\right.
\end{split}
\end{align}

\par
Under the above parameter configuration, Fig.~\ref{figure10} shows the influences of antenna pattern and half-power beamwidth on $6-$localizability probability. When the half-power beamwidth is larger, the beam is wider, and the interference signal from LEO satellites is stronger. Therefore, the $6-$localizability probability decreases with the half-power beamwidth. Compared to half-power beamwidth, the effect of the antenna pattern on the localizability probability is much smaller. Moreover, the analytical results are greater than the simulated ones, and this phenomenon is also observed in Fig.~\ref{figure6} and Fig.~\ref{figure8}. The main reason for this phenomenon is the approximation of interference. We only consider the impact of the closest interfering satellite on the localizability, while other interference sources are ignored. This leads to an underestimation of interference and an overestimation of localizability availability. However, the deviation is only about $0.3\%$, which is within an acceptable range.

\par
Finally, we point out the current limitations of beam gain modeling for further research and improvement. Firstly, the antenna patterns presented in Table~\ref{table1} are relatively basic, and not common. The antenna patterns adopted in current LEO satellite constellations are carefully designed, dynamic, and adaptive. Additionally, these four antenna patterns are all circularly symmetric, whereas in most scenarios, they are functions of two angular variables, namely the azimuth angle and the elevation angle relative to the beam's steering direction \cite{kim2023downlink}. Establishing a more meticulously designed beam gain model related to the satellite's latitude and longitude, along with developing an analytical framework based on it, represents a potential area for future research.

\subsection{Positioning Error}
The preceding analysis focuses on the unique challenges faced by stochastic geometry, the establishment of satellite spatial distribution and channel models, as well as the setup of satellite-enabled positioning problems. However, from the perspective of metrics, we are still focused on fundamental positioning-related metrics such as availability and localizability, without delving into performance metrics related to positioning accuracy. Undoubtedly, in comparison to localizability, positioning accuracy receives more widespread attention and research interest. An increase in localizability probability implies that more satellites will participate in the positioning process, thereby enhancing positioning accuracy. However, establishing a direct mapping relationship between localizability probability and positioning accuracy is challenging, and similar conclusions have yet to be reached.

\par
Fortunately, satellite-enabled positioning falls under geometric positioning, making positioning error suitable for analysis within the SG framework. Referring to research on terrestrial positioning systems based on stochastic geometry, two potential directions can be foreseen in the analysis of satellite-enabled positioning errors. The first direction involves the positioning error evaluation of specific geometric localization algorithms, such as the time difference of arrival algorithm. Along this research path, the core investigation may encompass estimating the probabilistic fuzzy region, also known as the circular error probable region \cite{lewis2007effects}.

\par
The second direction involves researching topics related to positioning error analysis but is not confined to the performance metrics of specific positioning algorithms, aligning more closely with the principles of stochastic geometry. The variance of SINR and Cram$\acute{e}$r-Rao lower bound (CRLB) are two typical metrics that meet the above condition. The variance of SINR reflects the fluctuation of signal quality. A larger SINR variance typically indicates a higher fluctuation in signal quality, which can lead to increased positioning errors. The CRLB can be used to estimate the theoretical lower bound of errors for all positioning algorithms. So far, the latter has been utilized in the performance evaluation of ground positioning networks under a stochastic geometric framework \cite{christopher2018statistical}.

\par
Finally, we discuss the challenges of transitioning from localizability to positioning error analysis. Stochastic geometry typically characterizes channel randomness using statistical models. Therefore, statistical models like the SR fading model usually focus on describing the long-term impact of fading and shadowing effects on signal strength. Moreover, localizability probability is a long-term metric that considers random variations in satellite positions that typically occur over several hours. However, factors such as ionospheric scintillation can cause instantaneous fluctuations in signal phase and amplitude, leading to inaccuracies in signal arrival time measurements by the receiver. Current statistical models have not yet accounted for this specific impact of ionospheric scintillation. Therefore, when dealing with critical factors affecting positioning accuracy like ionospheric scintillation, it is essential to not only consider its long-term effects through statistical models but also focus on its instantaneous impacts on positioning errors.

\subsection{SG-Based Models}
This section discusses recent advancements in point processes, a comparison between the CPP and DSBPP, and the implementation of DSBPP. In recent studies, point processes based on spherical stochastic geometry can be classified into non-orbital models \cite{yim2024modeling}, stochastic-orbital models \cite{choi2024analysis}, and fixed-orbital models \cite{sun2024performance}. The modeling accuracy of these three types of models increases gradually, but the complexity of deriving analytical expressions also correspondingly increases. When analyzing the time-correlated performance metrics of continuously moving satellites, the non-orbital model is no longer applicable. However, in scenarios involving complex topology analysis, such as routing, the strong tractability of the non-orbital model is essential. The stochastic-orbital model typically appears as a compromise choice, as in the context of this paper. 

\par
A detailed comparison of the strengths and weaknesses of DSBPP and CPP, both of which belong to stochastic orbital models, is also intriguing. CPP possesses many interesting mathematical characteristics, such as void probability and probability generating functional (PGFL), due to its Poisson randomness in quantity. These features play a crucial role in analyzing random variables like availability and interference, features that DSBPP lacks. Therefore, for a larger-scale LEO constellation like Starlink-1584, placed in $72$ orbital planes with $22$ satellites each, CPP is more tractable than DSBPP. Conversely, for a GPS MEO constellation composed of $6$ orbital planes with $4$ satellites each, assuming satellites on each orbital plane follow a Poisson distribution with a mean of $4$ rather than being fixed at $4$, would result in significant modeling differences. In such a scenario, DSBPP is preferred for modeling. 

\par
Finally, we present the implementation process of DSBPP in the form of MATLAB pseudo code. The following algorithm takes the number of orbits, the number of satellites on an orbit, and the distance between satellite and Earth center as inputs, and provides satellite positions as output. 

\begin{algorithm}[!ht] 
    \caption{DSBPP Generation Algorithm.}
    \begin{algorithmic} [1]
    \STATE \textbf{Input}: $N_o, N_M, R_M$. 

    \STATE \textbf{Initiate}: ${\mathrm{\textit{Pos}}} = {\mathrm{zeros}} (3, N_o N_M)$, and ${\mathrm{\textit{Index}}} = 1$.

    \STATE ${\mathrm{\textit{Inc}}}={\mathrm{acos}}(1-2*{\mathrm{rand}}(1,N_o))/\pi*180$.

    \STATE ${\mathrm{\textit{Azi}}}={\mathrm{rand}}(N_o,1)*360$.

    \FOR{$n = 1 : N_o$}
    \STATE ${\mathrm{\textit{Ang}}} = {\mathrm{rand}}(N_M, 1) * 360$.
    \FOR{$m = 1 : N_M$}

    \STATE $\textit{x} = R_M \, {\mathrm{cos}}( {\mathrm{\textit{Ang}}}(m) )$.

    \STATE $\textit{y} = R_M \,  {\mathrm{sin}}( {\mathrm{\textit{Ang}}}(m) )$.

    \STATE $\textit{z} = 0$.

    \STATE $\textit{Rx} = [1, 0, 0; 0, {\mathrm{cosd}}( {\mathrm{\textit{Inc}}} (n)), - {\mathrm{sind}}( {\mathrm{\textit{Inc}}} (n));0, \ \ $  ${\mathrm{sind}}({\mathrm{\textit{Inc}}} (n)), {\mathrm{cosd}}({\mathrm{\textit{Inc}}} (n))]$.
    
    \STATE $\textit{Rz} = [{\mathrm{cosd}}({\mathrm{\textit{Azi}}}(n)), -{\mathrm{sind}}({\mathrm{\textit{Azi}}}(n)), 0; \ \ \ \ \ \ $ ${\mathrm{sind}}({\mathrm{\textit{Azi}}}(n)),  {\mathrm{cosd}}({\mathrm{\textit{Azi}}}(n)), 0; 0, 0, 1]$.

    \STATE $\mathrm{\textit{Loc}} = \textit{Rz} * (\textit{Rx} * [\textit{x}; \textit{y}; \textit{z}]) $.

    \STATE ${\mathrm{\textit{Pos}}} (:, {\mathrm{\textit{Index}}}) = {\mathrm{\textit{Loc}}}$.
    \STATE ${\mathrm{\textit{Index}}} = {\mathrm{\textit{Index}}} + 1$.
    \ENDFOR
    \ENDFOR
    
    \STATE \textbf{Output}: \textit{Pos}.
    \end{algorithmic}
\end{algorithm}

Considering that the primary computational complexity of the algorithm stems from matrix multiplication, we define the complexity unit as one execution of step (12). As a result, the overall complexity of the algorithm is $\mathcal{O} (N_o N_M)$.

\section{Conclusion}\label{section5}
In this paper, we developed an SG-based theoretical framework for the availability and localizability analysis of hybrid LEO/MEO satellite systems. 
We also validated the accuracy of our analysis via numerical simulations, which also demonstrate the availability and localizability performance of LEO, MEO, and hybrid LEO/MEO constellations. 
Additionally, the analysis allows us to create heatmaps to explore the impact of satellite numbers on the above metrics. Finally, we demonstrated the effects of the beam center angle and antenna pattern on network positioning. For future research directions, LEO satellites are expected to take on an increasing role in positioning services with the ongoing improvement of positioning technology. Furthermore, multiple LEO constellations may collaboratively participate in positioning. Therefore, it is necessary to develop a more general association strategy for a hybrid constellation that includes multiple layers of LEO constellations and design a performance analysis framework based on this strategy within the SG framework.

\appendices
\section{Proof of Lemma~\ref{lemma1}}\label{app:lemma1}
In this proof, we study the upper limit of the central angle through the triangle formed by the Earth's center, satellite, and target. When the satellite is located at the horizon, this triangle forms a right-angled triangle, and $\theta_{Q,1}^{\max} = \arccos\left( {R_{\oplus}} / {R_Q} \right)$ in this case.

\par
When the target is exactly located at the boundary of the satellite's main lobe beamwidth, the following equation can be obtained from the Law of Sines:
\begin{equation}
    \frac{R_{\oplus}}{\sin\left( \frac{\varphi_Q}{2} \right)} = \frac{R_Q}{\sin\left( \pi -  \frac{\varphi_Q}{2} - \theta_{Q,2}^{\max} \right)}.
\end{equation}

When $0 < \pi -  \frac{\varphi_Q}{2} - \theta_{Q,2}^{\max} < \pi$,  there are two mappings from the sine value to the angle. When $\pi -  \frac{\varphi_Q}{2} - \theta_{Q,2}^{\max} \leq \frac{\pi}{2}$ is an acute angle or a right angle, $\theta_{Q,2}^{\max} = \arccos\left( {R_{\oplus}} / {R_Q} \right)$. (ii) Otherwise, when $\pi -  \frac{\varphi_Q}{2} - \theta_{Q,2}^{\max} > \frac{\pi}{2}$, 
\begin{equation}
\begin{split}
    \theta_{Q,2}^{\max} & = \pi - \frac{\varphi_Q}{2} - \left( \pi - \arcsin\left( \frac{R_L \sin\left( {\varphi_Q}/{2} \right)}{R_{\oplus}} \right) \right) \\
    & = \arcsin\left( \frac{R_L \sin\left( {\varphi_Q}/{2} \right)}{R_{\oplus}} \right) - \frac{\varphi_Q}{2}.
\end{split}
\end{equation}
It is not hard to prove that $\theta_{Q,2}^{\max} \leq \theta_{Q,1}^{\max}$ is always true.

\section{Proof of Lemma~\ref{lemma2}}\label{app:lemma2}

To start with, we derive the probability of a single LEO satellite located in the spherical cap with a distance less than or equal to $\theta_L^{\max}$ from the ground target. From the definition of BPP \cite{ok-1},
\begin{equation}\label{AppC-1}
\begin{split}
    & \mathbb{P}\left[ \theta \leq \theta_L^{\max} \right] = \frac{\mathcal{A}\left( \mathcal{S}(\theta_L^{\max}) \right)}{\mathcal{A}\left( \mathcal{S}(\pi)\right)} \\
    & = \frac{ 2\pi R_L^2 (1-\cos\theta_L^{\max})}{4\pi R_L^2} = \frac{ 1 - \cos\theta_L^{\max} }{2},
\end{split}
\end{equation}
where $\mathcal{S}(\theta)$ represents the spherical cap with a central angle of $2\theta$ and $\mathcal{A}\left( \mathcal{S}(\theta) \right)$ is the area measure of $\mathcal{S}(\theta)$. Considering that each LEO satellite located within the spherical cap $\mathcal{S}(\theta_L^{\max})$ is an independent event, the $K-$availability probability of LEO satellite networks can be obtained through the CDF of the binomial distribution:
\begin{equation}
\begin{split}
    P_L^A(K) & = 1 - \sum_{k=0}^{K-1} \mathbbm{P} \left[ \mathcal{N} \left( \mathcal{S}(\theta_L^{\max}) \right) = k \right] \\
    & = 1 - \sum_{k=0}^{K-1} \binom{N_L}{k} \left( \frac{ 1 - \cos\theta_L^{\max} }{2} \right)^k \\
    & \times \left( 1 - \frac{ 1 - \cos\theta_L^{\max} }{2} \right)^{N_L-k},
\end{split} 
\end{equation}
where $\mathcal{N} \left( \mathcal{S}(\theta_L^{\max}) \right)$ counts the number of LEO satellites in the sperical cap $\mathcal{S}(\theta_L^{\max})$.

\section{Proof of Lemma~\ref{lemma3}}\label{app:lemma3}
The proof of Lemma~\ref{lemma3} is divided into two steps: (i) Derive the critical condition for the existence of available satellites; (ii) Derive the maximum central angle $\theta_{c,M}^{\max}(\theta_{\perp},d_M^{\max})$.

\par
\begin{figure}[ht]
	\centering
    \includegraphics[width = 0.7\linewidth]{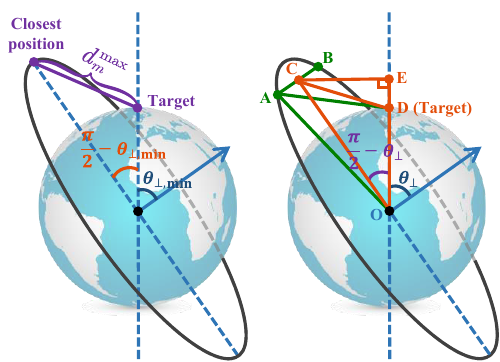}
	\caption{Illustration of the critical condition and maximum central angle.}
	\label{Appen_A}
\end{figure}

If the normal vector of the orbit is perpendicular to the typical target ($\theta_{\perp}=\frac{\pi}{2}$), the satellite moving trajectory will pass the overhead of the target. In this case, the typical target has the maximum satellite availability probability. As $\theta_{\perp}$ decreases, the orbit will progressively deviate from the target's overhead. To ensure there exist available satellites for the target, 
\begin{equation}
    \left| \theta_{\perp} - \frac{\pi}{2} \right| \leq \theta_{\perp,\min}
\end{equation}
needs to be satisfied. When   $\theta_{\perp}=\theta_{\perp,\min}$, the closest position on the orbit has a distance $d_{\max}$ to the target. As shown at the left part of Fig.~\ref{Appen_A}, we can present $\theta_{\perp,\min}$ by the law of cosines, 
\begin{equation}
    \frac{\pi}{2} - \theta_{\perp,\min} = \arccos \left( \frac{R_s^2 + R_{\oplus}^2 - (d_M^{\max})^2}{2R_s R_{\oplus}} \right).
\end{equation}
Therefore, the critical condition for the existence of available satellites is
\begin{equation}
    \left| \theta_{\perp} - \frac{\pi}{2} \right| \leq \arcsin \left( \frac{R_s^2 + R_{\oplus}^2 - (d_M^{\max})^2}{2R_s R_{\oplus}} \right).
\end{equation}

\par
Next, we derive the analytical expression for the maximum central angle. As shown in the right part of Fig.~\ref{Appen_A}, we denote the center of the Earth as $O$ and the position of the target as $D$. $A$ and $B$ are the critical positions in the orbit where satellites can maintain communication with the target. From the definition, we have $\overline{AD} = \overline{BD} = d_M^{\max}$, where $\overline{AD}$ is the length of the line segment $AD$. We draw a perpendicular from the midpoint $C$ of line segment $AB$ to the line $OD$, and the point of intersection is denoted as $E$. By the geometric relationship in right triangles $\triangle COE$ and $\triangle CDE$, it can be known that: 
\begin{align}
\begin{split}
    \left\{
 	\begin{array}{ll}
    & \! \! \! \! \! \overline{CE} = \overline{OC} \cos\theta_{\perp}, \\
    & \! \! \! \! \! \overline{OE} = \overline{OC} \sin\theta_{\perp} = \overline{DE} + R_{\oplus}, \\
    & \! \! \! \! \! \overline{DC}^2 = \overline{DE}^2 + \overline{CE}^2.
	\end{array}
	\right.
\end{split}
\end{align}
Thus, we get
\begin{equation}\label{AppA_1}
    \overline{DC}^2 = \overline{OC}^2 + R_{\oplus}^2 - 2\overline{OC} \sin\theta_{\perp} R_{\oplus}.
\end{equation}
By the geometric relationship in right triangles $\triangle AOC$ and $\triangle ADC$, it can be known that:
\begin{equation}\label{AppA_2}
\begin{split}
    \overline{DC}^2 = \overline{AD}^2 + \overline{OC}^2 - \overline{AO}^2 = d_{\max}^2 + \overline{OC}^2 - R_s^2.
\end{split}
\end{equation}
By combining (\ref{AppA_1}) and (\ref{AppA_2}), we get
\begin{equation}
    \overline{OC} = \frac{R_{\oplus}^2 + R_s^2 - (d_M^{\max})^2}{2R_{\oplus}\sin\theta_{\perp}}.
\end{equation}
Then, the maximum central angle for two available satellites in the orbit can be derived,
\begin{equation}
\begin{split}
    & \theta_{c,M}^{\max}(\theta_{\perp},d_M^{\max}) = 2 \arccos \left( \frac{\overline{OC}}{\overline{OA}} \right) = 2 \arccos \left( \frac{R_{\oplus}^2 + R_s^2 - d_{\max}^2}{2R_{\oplus} R_s \sin\theta_{\perp}} \right).
    \end{split}
\end{equation}
Please note that the above letters ($O,A \sim F$) and symbols (such as $\overline{AB}$) are only defined for the convenience of theorem proof. Their meanings are different from those of letters and symbols in other parts of the article.

\section{Proof of Lemma~\ref{lemma4}}\label{app:lemma4}
In this proof, we derive the probability that the distance from a single MEO satellite to the target is less than $\theta_M^{\max}$. Given that an MEO satellite is located on an orbit with inclination angle $\theta_{\perp}$, the probability that this MEO satellite is available to the target is ${\theta_{c,M}^{\max}(\theta_{\perp},d_M^{\max})}/{(2\pi)}$. Therefore, the probability that a single MEO satellite is
available for the target is
\begin{equation}\label{AppD-1}
\begin{split}
    & p_{M,1}^A = \mathbbm{E}_{\theta_{\perp}} \left[ \frac{\theta_{c,M}^{\max}(\theta_{\perp},d_M^{\max})}{2\pi} \right] = \int_{\frac{\pi}{2} - \arccos \left( \frac{R_\oplus^2 + R_M^2 - (d_M^{\max})^2}{2 R_\oplus R_M} \right)}^{ \frac{\pi}{2} + \arccos \left( \frac{R_\oplus^2 + R_M^2 - (d_M^{\max})^2}{2 R_\oplus R_M} \right) } \frac{ \sin\theta}{2} \frac{\theta_{c,M}^{\max} \left( \theta,d_M^{\max} \right)}{2\pi} {\mathrm{d}} \theta.
\end{split}
\end{equation}
Note that the upper and lower bounds of integral in (\ref{AppD-1}) come from the fact that  $\theta_{c,M}^{\max}(\theta_{\perp},d_M^{\max}) \neq 0$ when 
\begin{equation}
    \left| \theta_\perp - \frac{\pi}{2} \right| \leq \arccos \left( \frac{R_\oplus^2 + R_M^2 - (d_M^{\max})^2}{2 R_\oplus R_M} \right).
\end{equation}

\par
Similarly, the probability of exactly $k$ available MEO satellites $p_M^A (k)$ can also be obtained using the CDF of the binomial distribution:
\begin{equation}
    P_M^A (K) = 1 - \sum_{k=0}^{K-1} \binom{N_M N_O}{k} \left( p_{M,1}^A \right)^k \left( 1 - p_{M,1}^A \right)^{N_M N_O-k}.
\end{equation}

\section{Proof of Theorem~\ref{theorem1}}\label{app:theorem1}
The $K$-availiability probability for the hybrid LEO/MEO satellite networks can be categorized into the following two conditions:
\begin{itemize}
\item When the number of available MEO satellites is greater than $K$, the $K$-availability is met regardless of the availability of LEO satellites. This condition happens with probability
\begin{equation}
    P_{{\mathrm{all}},1}^A(K) =  \sum_{k=K}^{\min \{ K , N_M^{\max} \}} p_M^A (k),
\end{equation}
where $p_M^A (k)$ can be obtained by probability mass function (PMF) of the binomial distribution,
\begin{equation}
p_M^A (k) = \binom{N_M N_O}{k} \left( p_{M,1}^A \right)^k \left( 1 - p_{M,1}^A \right)^{N_M N_O-k}.
\end{equation}
\item If exactly $k < K$ MEO satellites are available, the probability of achieving $k-$availability in a hybrid constellation is equivalent to the probability of having at least $K - k$ available LEO satellites. By traversing the available quantities of MEO satellites $k$, this condition happens with probability
\begin{equation}
    P_{{\mathrm{all}},2}^A(K) =  \sum_{k=0}^{\min\{K-1,N_M^{\max}-1\}} P_L^A ( K - k ) \, p_M^A (k).
\end{equation}
\end{itemize}
The final $K$-availability probability of the hybrid constellation can be obtained by
\begin{equation}
    P_{\mathrm{all}}^A(K) = P_{{\mathrm{all}},1}^A(K) + P_{{\mathrm{all}},2}^A(K).
\end{equation}

\section{Proof of Lemma~\ref{lemma5}}\label{app:lemma5}
According to the CDF of the binomial distribution, the CDF of $k-$LEO contact angle distribution is derived as,
\begin{equation}\label{AppB-1}
\begin{split}
    & F_{\theta_L}^{(k)} (\theta) =  1 - \sum_{n=0}^{k-1} \mathbbm{P} \left[ \mathcal{N} \left( \mathcal{S} \left( \theta \right) \right) = n \right] \\
    & = 1 - \sum_{n=0}^{k-1} \binom{N_L}{n} \left( \frac{1-\cos\theta}{2}\right)^n \left( \frac{1+\cos\theta}{2}\right)^{N_L-n}.
\end{split}
\end{equation}
where $\mathcal{S}(\theta)$ denotes the spherical cap with a central angle of $2\theta$, and $\mathcal{N} \left( \mathcal{S}(\theta) \right)$ counts the number of LEO satellites $\mathcal{S}(\theta)$. Take the derivative of $F_{\theta_L}^{(k)} (\theta)$ yields the PDF of the contact angle distribution, which is given in (\ref{AppF-2}) at the top of the next page.
\begin{table*}
\begin{equation}\label{AppF-2}
\begin{split}
    & f_{\theta_L}^{(k)} (\theta) = 1 - \frac{1}{2^{N_L}} \sum_{n=0}^{k-1} \binom{N_L}{n} \times \bigg( \left( 1-\cos\theta \right)^n \frac{\mathrm{d} \left( 1+\cos\theta \right)^{N_L-n}}{\mathrm{d} \theta} + \left( 1+\cos\theta \right)^{N_L-n} \frac{\mathrm{d} \left( 1-\cos\theta \right)^n}{\mathrm{d} \theta} \bigg) \\
    & = 1 - \frac{\sin\theta}{2^{N_L}} \sum_{n=0}^{k-1} \binom{N_L}{n} \bigg( n \left( 1+\cos\theta \right)^{N_L-n} \left( 1-\cos\theta \right)^{n-1} - \left( N_L - n \right) \left( 1-\cos\theta \right)^n \left( 1+\cos\theta \right)^{N_L-n-1} \bigg) \\
    & = 1 - \frac{\sin\theta}{2^{N_L}} \sum_{n=0}^{k-1} \binom{N_L}{n} \left( 1- \cos \theta \right)^n \left( 1+\cos\theta \right)^{N_L-n} \left( \frac{n}{1-\cos\theta} - \frac{N_L - n}{1+\cos\theta} \right).
\end{split}
\end{equation}
\hrule
\end{table*}

\section{Proof of Lemma~\ref{lemma6}}\label{app:lemma6}
To start with, we derive the conditional MEO contact angle distribution for a given inclination angle,
\begin{equation}
\begin{split}
    & F_{\theta_M} \left( \theta \, | \, \theta_{\perp} \right) = \mathbbm{P} \left[ \theta_M \leq \theta \, | \, \theta_{\perp} \right] \\
    & = \frac{1}{2\pi} \, \theta_{c,M}^{\max} \left( \theta_{\perp}, \sqrt{R_M^2 + R_{\oplus}^2 - 2R_M R_{\oplus}\cos\theta } \right) \\
    & = \frac{1}{\pi} \arccos \left( \frac{\cos\theta}{\sin\theta_{\perp}} \right), \ \theta \leq \theta_M^{\max},
\end{split}
\end{equation}
where $\theta_{c,M}^{\max} \left( \theta_{\perp},l(\theta) \right)$ is derived in Lemma~\ref{lemma3}, and $l(\theta) = \sqrt{R_M^2 + R_{\oplus}^2 - 2R_M R_{\oplus}\cos\theta }$ denotes the Euclidean distance between the MEO satellite and target given that the central angle between them is $\theta$. 

\par
On an orbit with an inclination angle of $\theta_{\perp}$, the minimum contact angle from an MEO satellite to the target is $\left| \frac{\pi}{2} - \theta_{\perp} \right|$. If there exist an MEO satellite with MEO contact angle $\theta_M \leq \theta$, the range of inclination angle should satisfy $\left| \frac{\pi}{2} - \theta_{\perp} \right| \leq \theta$. Therefore, the unconditional MEO contact angle distribution can be obtained by
\begin{equation}
\begin{split}
    F_{\theta_M} (\theta) & = \int_{\frac{\pi}{2}-\theta}^{\frac{\pi}{2}+\theta} \frac{\sin\theta_{\perp}}{2} \frac{1}{\pi} \arccos \left( \frac{\cos\theta}{\sin\theta_{\perp}} \right) {\mathrm{d}} \theta_{\perp} \\
    & = \int_{\frac{\pi}{2}-\theta}^{\frac{\pi}{2}} \frac{\sin\theta_{\perp}}{\pi} \arccos \left( \frac{\cos\theta}{\sin\theta_{\perp}} \right) {\mathrm{d}} \theta_{\perp}.
\end{split}
\end{equation}

\par
According to the Leibniz integral rule, the PDF of MEO contact angle distribution can be obtained by taking the derivative of the CDF,
\begin{equation}
\begin{split}
    & f_{\theta_M} (\theta) = \int_{\frac{\pi}{2}-\theta}^{\frac{\pi}{2}} \frac{\sin\theta_{\perp}}{\pi} \frac{{\mathrm{d}}}{{\mathrm{d}}\theta} \left( \arccos \left( \frac{\cos\theta}{\sin\theta_{\perp}} \right) \right) {\mathrm{d}} \theta_{\perp} \\
    & - \frac{{\mathrm{d}}\left( \frac{\pi}{2} - \theta \right)}{{\mathrm{d}}\theta} \times \frac{\sin\left( \frac{\pi}{2} - \theta \right)}{\pi} \arccos \left( \frac{\cos\theta}{\sin\left( \frac{\pi}{2} - \theta \right)} \right) \\
    & \overset{(a)}{=} \frac{1}{\pi} \int_{\frac{\pi}{2}-\theta}^{\frac{\pi}{2}} \sin\theta_{\perp} \frac{{\mathrm{d}}}{{\mathrm{d}}\theta} \left( \arccos \left( \frac{\cos\theta}{\sin\theta_{\perp}} \right) \right) {\mathrm{d}} \theta_{\perp} \\
    & = \frac{1}{\pi} \int_{\frac{\pi}{2}-\theta}^{\frac{\pi}{2}} \sin\theta_{\perp} \frac{\sin\theta}{\sqrt{1 - \left( \frac{\cos\theta}{\sin\theta_{\perp}}   \right)^2}} {\mathrm{d}} \theta_{\perp},
\end{split}
\end{equation}
where step $(a)$ comes from
\begin{equation}
    \arccos \left( \frac{\cos\theta}{\sin\left( \frac{\pi}{2} - \theta \right)} \right) = \arccos \left( \frac{\cos\theta}{\cos\theta} \right) = 0. 
\end{equation}

\section{Proof of Lemma~\ref{lemma7}}\label{app:lemma7}
The proof is divided into the following steps:
\begin{itemize}
    \item Introduce the concept of effective receiving beam range.
    \item Establish the relationship between the dome angle and the central angle.
    \item Analyze the two most likely conditions and their probabilities for the distribution of interfering satellites.
    \item Derive the PDF of the interference power distribution.
\end{itemize}

Note that The dome angle between A and B is the angle created by the line connecting A to the ground target and the line connecting B to the ground target. Taking the Gaussian antenna pattern as an example, When $\varphi^r > 3\varphi_{\mathrm{3dB}}$, $G^r(\varphi^r) < 2^{-9} \, G_L^m$. Therefore, power from interfering satellites with dome angle beyond $\varphi_{\max} = 3\varphi_{\mathrm{3dB}}$ at the target can be neglected due to the excessively low receiving antenna gain. For other patterns in Table~\ref{table1}, we consider $\varphi_{\max} = \varphi_{\mathrm{3dB}}$ for Flat-top antenna, $\varphi_{\max} = \frac{3}{N_a}$ for Sinc antenna (including main lobe and the first side lobe), and $\varphi_{\max} = \frac{1}{N_a}$ for Cosine antenna. 

\par
Then, relationship between central angle $\theta_d^{\max}$ and the corresponding dome angle $\varphi_{\max}$ can be given through the sine rule \cite{Al-1},
\begin{equation}
    \varphi_{\max} = \cot^{-1} \left( \cot\theta_d^{\max} - \frac{R_{\oplus}}{R_L} \sqrt{1+\cot^2\theta_d^{\max}} \right).
\end{equation}
By applying the inverse operation, the result is as follows:
\begin{equation}
\begin{split}
    & \theta_d^{\max} = \cot^{-1} \Bigg( \frac{R_L^2}{R_L^2 - R_{\oplus}^2} \bigg( \cot \varphi_{\max} \\
    & + 
    \sqrt{(R_{\oplus}/R_L)^2 \left(1 + \cot^2\varphi_{\max} - (R_{\oplus}/R_L)^2 \right) } \bigg) \Bigg). 
\end{split}
\end{equation}

\par
To reduce computational complexity, we consider the two most likely conditions to approximate the interference power distribution: there is zero or one available LEO satellite within the effective receiving beam range. Typically, the number of satellites in MEO constellations is much smaller than that in LEO constellations. Therefore, under the narrow beam assumption, we omit the availability of MEO satellites. Similar to the proof of Lemma~\ref{lemma2}, the probability of having no satellites in the effective receiving beam range is 
\begin{equation} 
\begin{split}
    & P_{\varphi}^{(0)}(\theta_d^{\max})  = \mathbbm{P} \left[ \mathcal{N} \left( \mathcal{S}(\theta_d^{\max}) \right) = 0 \right] \\
    & = \left( 1 - \frac{ 2 \pi R_L^2 (1-\cos\theta_d^{\max})  }{4 \pi R_L^2} \right)^{N_L} = \left( \frac{1 + \cos\theta_d^{\max}}{2}  \right)^{N_L}.
\end{split} 
\end{equation}
The above deduction is based on the spherical cap approximation of the beam shape under the narrow-beam assumption. The probability of having one satellite can be approximately considered as $P_{\varphi}^{(1)} \approx 1 - P_{\varphi}^{(0)}$.

\par
Finally, when the central angle of the effective receiving beam range is $\theta_d^{\max}$, the interference power $I(\theta_d^{\max}) = 0$ with probability $P_{\varphi}^{(0)}(\theta_d^{\max})$. Given that there is one satellite within the effective receiving beam range, the CDF of the central angle between the associated satellite and the interfering satellite is
\begin{equation}
    F_{\theta_{I}} (\theta) = \frac{2\pi R_L^2 (1-\cos\theta)}{2\pi R_L^2 (1-\cos\theta_d^{\max})} = \frac{1-\cos\theta}{1-\cos\theta_d^{\max}},
\end{equation}
where $0 \leq \theta \leq \theta_d^{\max}$. Given that the central angle between the associated satellite and the interfering satellite is $\theta_{I}$, the interference power can be approximately expressed as
\begin{equation}
\begin{split}
    & I_{|\theta_I} = \frac{\rho_L^t G_L^t  \zeta \left(\frac{\lambda_L}{4\pi}\right)^2 W}{R_L^2 + R_{\oplus}^2 - 2 R_L R_{\oplus} \cos\theta_L} \\
    & \times G^r \left( \cot^{-1} \left( \cot\theta_I - \frac{R_{\oplus}}{R_L} \sqrt{1+\cot^2\theta_I} \right) \right),
\end{split}
\end{equation}
where $\theta_L$ is the central angle between the associated satellite and the target. Therefore, the CDF of the interference power distribution is 
\begin{equation}\label{AppH-1}
\begin{split}
    & F_{I} (I) = \mathbbm{E}_{\theta_I}\left[ I_{|\theta_I} < I \right] \\
    & = \int_{0}^{\theta_d^{\max}} \frac{\sin\theta_I}{1-\cos \theta_d^{\max}} \, F_W \Bigg( I \frac{R_L^2 + R_{\oplus}^2 - 2 R_L R_{\oplus} \cos\theta_L}{\rho_L^t G_L^t  \zeta \left( \frac{\lambda_L}{4\pi}\right)^2} \\
    & \times \left( G^r \left( \cot^{-1} \left( \cot\theta_I - \frac{R_{\oplus}}{R_L} \sqrt{1+\cot^2\theta_I} \right) \right) \right)^{-1} \Bigg) {\mathrm{d}} \theta_I. 
\end{split}
\end{equation}
The PDF of interference power distribution can be obtained by taking the derivative of (\ref{AppH-1}) with respect to $I$.

\section{Proof of Lemma~\ref{lemma8}}\label{app:lemma8}
First, we start the derivation from the localizability probability for the $k$-nearest satellites to the target:
\begin{equation}
\begin{split}
    & p_L^C (k) = \mathbbm{E}_{I,\theta_L} \bigg[ \mathbbm{P} \bigg[ \left(\frac{\lambda_L}{4\pi}\right)^2 \frac{1}{\left(I+\sigma_L^2\right)} \\
    & \times \frac{\rho_L^t G_L^t  G_L^m \zeta W}{ \left( R_L^2 + R_{\oplus}^2 - 2R_L R_{\oplus} \cos\theta_L \right) } > \gamma_L \bigg] \bigg] \\
    & = \int_0^{\theta_L^{\max}} f_{\theta_L}^{(k)}(\theta) \int_0^{\infty} f_I(I) \Bigg( 1 - F_W \Bigg( \gamma_L \left(  \frac{4\pi}{\lambda_L} \right)^2 \\
    & \times \frac{\left(I+\sigma_L^2\right) \left( R_L^2 + R_{\oplus}^2 - 2R_L R_{\oplus} \cos\theta_L \right) }{\rho_L^t G_L^t G_L^m \zeta} \Bigg) \Bigg) {\mathrm{d}}I {\mathrm{d}}\theta_L.
\end{split}
\end{equation}
From the definition, the $K$-localizability probability for the LEO satellite constellation is
\begin{equation}
    P_L^C (K) = \prod_{k=1}^K p_L^C (k).
\end{equation}

\bibliographystyle{IEEEtran}
\bibliography{references}

\begin{thebibliography}{10}
\providecommand{\url}[1]{#1}
\csname url@samestyle\endcsname
\providecommand{\newblock}{\relax}
\providecommand{\bibinfo}[2]{#2}
\providecommand{\BIBentrySTDinterwordspacing}{\spaceskip=0pt\relax}
\providecommand{\BIBentryALTinterwordstretchfactor}{4}
\providecommand{\BIBentryALTinterwordspacing}{\spaceskip=\fontdimen2\font plus
\BIBentryALTinterwordstretchfactor\fontdimen3\font minus \fontdimen4\font\relax}
\providecommand{\BIBforeignlanguage}[2]{{%
\expandafter\ifx\csname l@#1\endcsname\relax
\typeout{** WARNING: IEEEtran.bst: No hyphenation pattern has been}%
\typeout{** loaded for the language `#1'. Using the pattern for}%
\typeout{** the default language instead.}%
\else
\language=\csname l@#1\endcsname
\fi
#2}}
\providecommand{\BIBdecl}{\relax}
\BIBdecl

\bibitem{cai2015precise}
C.~Cai, Y.~Gao, L.~Pan, and J.~Zhu, ``Precise point positioning with quad-constellations: {GPS}, {BeiDou}, {GLONASS} and {G}alileo,'' \emph{Advances in space research}, vol.~56, no.~1, pp. 133--143, 2015.

\bibitem{ferre2022leo}
R.~M. Ferre, E.~S. Lohan, H.~Kuusniemi, J.~Praks, S.~Kaasalainen, C.~Pinell, and M.~Elsanhoury, ``Is {LEO}-based positioning with mega-constellations the answer for future equal access localization?'' \emph{IEEE Communications Magazine}, vol.~60, no.~6, pp. 40--46, 2022.

\bibitem{liao2023integration}
Y.~Liao, S.~Liu, X.~Hong, J.~Shi, and L.~Cheng, ``Integration of communication and navigation technologies toward {LEO}-enabled {6G} networks: {A} survey,'' \emph{Space: Science \& Technology}, vol.~3, p. 0092, 2023.

\bibitem{yue2022satellite}
F.~Yue, Z.~Cui, S.~Li, H.~Jing, S.~Zhang, and M.~Wang, ``A satellite augmentation system based on {LEO} mega-constellation,'' in \emph{Proc. IEEE International Conference on Artificial Intelligence, Information Processing and Cloud Computing (AIIPCC)}, 2022, pp. 221--225.

\bibitem{reid2018broadband}
T.~G. Reid, A.~M. Neish, T.~Walter, and P.~K. Enge, ``Broadband {LEO} constellations for navigation,'' \emph{NAVIGATION: Journal of the Institute of Navigation}, vol.~65, no.~2, pp. 205--220, 2018.

\bibitem{ferre2021comparison}
R.~M. Ferre and E.~S. Lohan, ``Comparison of {MEO}, {LEO}, and terrestrial {IoT} configurations in terms of gdop and achievable positioning accuracies,'' \emph{IEEE Journal of Radio Frequency Identification}, vol.~5, no.~3, pp. 287--299, 2021.

\bibitem{specht2015accuracy}
C.~Specht, M.~Mania, M.~Sk{\'o}ra, and M.~Specht, ``Accuracy of the gps positioning system in the context of increasing the number of satellites in the constellation,'' \emph{Polish Maritime Research}, vol.~22, no.~2, pp. 9--14, 2015.

\bibitem{raghu2016tracking}
N.~Raghu, B.~Kiran, and K.~Manjunatha, ``Tracking of {IRNSS}, {GPS} and hybrid satellites by using {IRNSS} receiver in {STK} simulation,'' in \emph{International Conference on Communication and Signal Processing (ICCSP)}.\hskip 1em plus 0.5em minus 0.4em\relax IEEE, 2016, pp. 0891--0896.

\bibitem{halevi2022asymptotic}
H.~Halevi, I.~Bergel, and Y.~Noam, ``Asymptotic performance of {TDOA} estimation using satellites,'' \emph{IEEE Transactions on Signal Processing}, vol.~70, pp. 2349--2361, 2022.

\bibitem{wei2023time}
Q.~Wei, X.~Chen, C.~Jiang, and Z.~Huang, ``Time-of-arrival estimation for integrated satellite navigation and communication signals,'' \emph{IEEE Transactions on Wireless Communications}, 2023.

\bibitem{chandrika2023spin}
V.~R. Chandrika, J.~Chen, L.~Lampe, G.~Vos, and S.~Dost, ``{SPIN}: {Synchronization} signal based positioning algorithm for {IoT} non-terrestrial networks,'' \emph{IEEE Internet of Things Journal}, 2023.

\bibitem{al2022next}
B.~Al~Homssi, A.~Al-Hourani, K.~Wang, P.~Conder, S.~Kandeepan, J.~Choi, B.~Allen, and B.~Moores, ``Next generation mega satellite networks for access equality: {O}pportunities, challenges, and performance,'' \emph{IEEE Communications Magazine}, vol.~60, no.~4, pp. 18--24, 2022.

\bibitem{wang2022ultra}
R.~Wang, M.~A. Kishk, and M.-S. Alouini, ``Ultra-dense {LEO} satellite-based communication systems: {A} novel modeling technique,'' \emph{IEEE Communications Magazine}, vol.~60, no.~4, pp. 25--31, 2022.

\bibitem{Al-2}
A.~Al-Hourani, ``Optimal satellite constellation altitude for maximal coverage,'' \emph{IEEE Wireless Communications Letters}, vol.~10, no.~7, pp. 1444--1448, 2021.

\bibitem{wang2025modeling}
R.~Wang, M.~A. Kishk, and M.-S. Alouini, ``Modeling and analysis of non-terrestrial networks by spherical stochastic geometry,'' 2025, available online: https://arxiv.org/abs/2503.13455.

\bibitem{talgat2020stochastic}
A.~Talgat, M.~A. Kishk, and M.-S. Alouini, ``Stochastic geometry-based analysis of {LEO} satellite communication systems,'' \emph{IEEE Communications Letters}, vol.~25, no.~8, pp. 2458--2462, 2021.

\bibitem{wang2022evaluating}
R.~Wang, M.~A. Kishk, and M.-S. Alouini, ``Evaluating the accuracy of stochastic geometry based models for {LEO} satellite networks analysis,'' \emph{IEEE Communications Letters}, vol.~26, no.~10, pp. 2440--2444, 2022.

\bibitem{ok-1}
N.~Okati, T.~Riihonen, D.~Korpi, I.~Angervuori, and R.~Wichman, ``Downlink coverage and rate analysis of low {E}arth orbit satellite constellations using stochastic geometry,'' \emph{IEEE Transactions on Communications}, vol.~68, no.~8, pp. 5120--5134, 2020.

\bibitem{choi2024novel}
C.-S. Choi and F.~Baccelli, ``A novel analytical model for leo and meo satellite networks based on cox point processes,'' \emph{IEEE Transactions on Communications}, 2024, early Access.

\bibitem{choi2024modeling}
C.-S. Choi, ``Modeling and analysis of downlink communications in a heterogeneous {LEO} satellite network,'' \emph{IEEE Transactions on Wireless Communications}, 2024, early access.

\bibitem{jung2023modeling}
D.-H. Jung, H.~Nam, J.~Choi, and D.~J. Love, ``Modeling and analysis of {GEO} satellite networks,'' \emph{\rm{available online: https://arxiv.org/abs/2312.15924}}.

\bibitem{schloemann2015localization}
J.~Schloemann, H.~S. Dhillon, and R.~M. Buehrer, ``Localization performance in cellular networks,'' in \emph{Proc. IEEE International Conference on Communication Workshop (ICCW)}, 2015, pp. 871--876.

\bibitem{christopher2018statistical}
E.~Christopher, H.~S. Dhillon, and R.~M. Buehrer, ``A statistical characterization of localization performance in wireless networks,'' \emph{IEEE Transactions on Wireless Communications}, vol.~17, no.~9, pp. 5841--5856, 2018.

\bibitem{schloemann2015toward}
J.~Schloemann, H.~S. Dhillon, and R.~M. Buehrer, ``Toward a tractable analysis of localization fundamentals in cellular networks,'' \emph{IEEE Transactions on Wireless Communications}, vol.~15, no.~3, pp. 1768--1782, 2015.

\bibitem{talgat2020nearest}
A.~Talgat, M.~A. Kishk, and M.-S. Alouini, ``Nearest neighbor and contact distance distribution for binomial point process on spherical surfaces,'' \emph{IEEE Communications Letters}, vol.~24, no.~12, pp. 2659--2663, 2020.

\bibitem{okati2022nonhomogeneous}
N.~Okati and T.~Riihonen, ``Nonhomogeneous stochastic geometry analysis of massive {LEO} communication constellations,'' \emph{IEEE Transactions on Communications}, vol.~70, no.~3, pp. 1848--1860, 2022.

\bibitem{wang2022stochastic}
R.~Wang, M.~A. Kishk, and M.-S. Alouini, ``Stochastic geometry-based low latency routing in massive {LEO} satellite networks,'' \emph{IEEE Transactions on Aerospace and Electronic Systems}, vol.~58, no.~5, pp. 3881--3894, 2022.

\bibitem{feller1991introduction}
W.~Feller, \emph{An introduction to probability theory and its applications, Volume 2}.\hskip 1em plus 0.5em minus 0.4em\relax John Wiley \& Sons, 1991, vol.~81.

\bibitem{abdi2003new}
A.~Abdi, W.~C. Lau, M.-S. Alouini, and M.~Kaveh, ``A new simple model for land mobile satellite channels: {F}irst-and second-order statistics,'' \emph{IEEE Transactions on Wireless Communications}, vol.~2, no.~3, pp. 519--528, 2003.

\bibitem{almagbile2010evaluating}
A.~Almagbile, J.~Wang, and W.~Ding, ``Evaluating the performances of adaptive {K}alman filter methods in {GPS/INS} integration,'' \emph{Journal of Global Positioning Systems}, vol.~9, no.~1, pp. 33--40, 2010.

\bibitem{singhal2023leo}
S.~Singhal, S.~K. Biswas, and S.~S. Ram, ``{LEO/MEO}-based multi-static passive radar detection performance analysis using stochastic geometry,'' in \emph{Radar Conference (RadarConf23)}.\hskip 1em plus 0.5em minus 0.4em\relax IEEE, 2023, pp. 1--6.

\bibitem{gagliardi2012satellite}
R.~M. Gagliardi, \emph{Satellite communications}.\hskip 1em plus 0.5em minus 0.4em\relax Springer Science \& Business Media, 2012.

\bibitem{balanis2015antenna}
C.~A. Balanis, \emph{Antenna theory: analysis and design}.\hskip 1em plus 0.5em minus 0.4em\relax John wiley \& sons, 2015.

\bibitem{yu2017coverage}
X.~Yu, J.~Zhang, M.~Haenggi, and K.~B. Letaief, ``Coverage analysis for millimeter wave networks: The impact of directional antenna arrays,'' \emph{IEEE Journal on Selected Areas in Communications}, vol.~35, no.~7, pp. 1498--1512, 2017.

\bibitem{kim2023downlink}
E.~Kim, I.~P. Roberts, and J.~G. Andrews, ``Downlink analysis and evaluation of multi-beam {LEO} satellite communication in shadowed {R}ician channels,'' \emph{IEEE Transactions on Vehicular Technology}, vol.~73, no.~2, pp. 2061--2075, 2024.

\bibitem{lewis2007effects}
J.~S. Lewis, J.~L. Rachlow, E.~O. Garton, and L.~A. Vierling, ``Effects of habitat on {GPS} collar performance: using data screening to reduce location error,'' \emph{Journal of applied ecology}, vol.~44, no.~3, pp. 663--671, 2007.

\bibitem{yim2024modeling}
J.~Yim, J.~Park, and N.~Lee, ``Modeling and coverage analysis of {K}-tier integrated satellite-terrestrial downlink networks,'' 2024, available online: https://arxiv.org/abs/2403.11096.

\bibitem{choi2024analysis}
C.-S. Choi, ``Analysis of a delay-tolerant data harvest architecture leveraging low earth orbit satellite networks,'' \emph{IEEE Journal on Selected Areas in Communications}, vol.~42, no.~5, pp. 1329--1343, 2024.

\bibitem{sun2024performance}
Y.~Sun and R.~Li, ``Performance analysis of satellite-terrestrial integrated radio access networks based on stochastic geometry,'' 2024, available online: https://arxiv.org/abs/2404.09506.

\bibitem{Al-1}
A.~Al-Hourani, ``An analytic approach for modeling the coverage performance of dense satellite networks,'' \emph{IEEE Wireless Communications Letters}, vol.~10, no.~4, pp. 897--901, 2021.

\end{thebibliography}

\end{document}